\documentclass[a4paper,11pt]{article}

\usepackage{jheppub} 

\usepackage{bm}
\usepackage{soul}

\usepackage[T1]{fontenc} 

\newcommand\Schrodinger {Schr\"{o}dinger }

\title{\boldmath Entanglement Entropy of a Scalar Field in Anti-de Sitter Space}

\author[a]{K. Boutivas,}
\author[a]{D. Katsinis,}
\author[a]{I. Papadimitriou,}
\author[a,b]{G. Pastras}
\author[a]{and N. Tetradis}

\affiliation[a]{Department of Physics, University of Athens, Zographou 157 84, Greece}
\affiliation[b]{Laboratory for Manufacturing Systems and Automation, Department of Mechanical Engineering and Aeronautics, University of Patras, Patra 26110, Greece}

\emailAdd{kboutivas@phys.uoa.gr}
\emailAdd{dkatsinis@phys.uoa.gr}
\emailAdd{ioannis.papadimitriou@phys.uoa.gr}
\emailAdd{pastras@lms.mech.upatras.gr}
\emailAdd{ntetrad@phys.uoa.gr}

\abstract{We study the entanglement entropy of a free massive scalar field at its ground state in $(3+1)$-dimensional AdS space in global coordinates. We consider spherical entangling surfaces centered at the origin of AdS. We determine the structure of the UV-divergent terms in the entanglement entropy and compute the numerical values of the respective coefficients. We confirm the connection between the coefficient of the logarithmic term and the conformal anomaly.}

\begin{document} 
\maketitle
\flushbottom

\section{Introduction}\label{sec:introduction}

The study of entanglement entropy of scalar fields in $3+1$ dimensions was initiated a long time ago \cite{Sorkin:1984kjy,Bombelli:1986rw,Srednicki:1993im}. The calculation by Srednicki for a massless scalar field in Minkowski space at its ground state \cite{Srednicki:1993im} produced the remarkable result that the entanglement entropy is dominated by an area-law term, similarly to the entropy of black holes. For a spherical entangling surface of radius $R$, the leading contribution to the entropy is proportional to $R^2/\epsilon^2$, where $\epsilon$ is a cutoff regulating the UV divergence that results from the strong entanglement of degrees of freedom very close to the entangling surface. The first subleading term is logarithmic, of the form $ \ln(R/\epsilon)$ with a universal coefficient equal to $-1/90$ \cite{Solodukhin:2008dh,Casini:2010kt,Casini:2009sr,Lohmayer:2009sq}. 

The extension of these results to curved backgrounds is very interesting, especially in the presence of horizons. In the context of the analysis, several non-trivial choices must be made that demand careful consideration. A fundamental choice concerns the state of the system, which should be selected to match the physical context of the calculation. When a horizon is present, this choice becomes more difficult, as the system's state behind it may be unknown. The choice of coordinates is also crucial in practice, as it determines the slices on which the entropy is computed and selects specific entangling surface geometries.

Extending Srednicki's result to curved backgrounds requires overcoming several technical challenges. The first one is connected to the presence of several scales. In flat space, the only intrinsic scale of the theory is the UV cutoff.\footnote{An IR cutoff is also present when the numerical calculation is carried out using a finite, discretized version of the theory. However, the goal of the numerical calculation is to obtain the entanglement entropy in the infinite-size limit, where the dependence of physical quantities on this scale ceases to exist.} On the other hand, a scalar field theory defined on a curved background, even in symmetric spaces such as de Sitter (dS) or anti-de Sitter (AdS), naturally contains at least one additional scale connected to the curvature. This fact introduces dependence on numerous dimensionless ratios of the various scales. Consequently, entanglement entropy may contain many more terms than it does in the case of a flat background. Moreover, the expected dependence of the entropy, not only on various invariants of the bulk geometry, but also on the extrinsic curvature of the entangling surface, makes calculations very challenging. Furthermore, there are many coordinate systems available in such backgrounds. The choice of coordinates influences the density of degrees of freedom in the discretized theory. While this should not impact the results for scheme-independent terms in the continuous limit, its effect on scheme-dependent terms must be analyzed carefully. Despite these challenges, concrete results can be obtained for curved spaces of high interest.

The entanglement entropy of a scalar field in de Sitter (dS) space was studied analytically by Maldacena and Pimentel in flat slicing, using the so-called planar coordinates employed in cosmology \cite{Maldacena:2012xp}. For a spherical entangling surface of proper area $A$, the entropy was parameterized as\footnote{We have modified slightly the definition of the parameters relative to \cite{Maldacena:2012xp}.}
\begin{equation}\label{eq:SEE_expansion_mald}
S_{\textrm{dS}}=\frac{c_1}{4\pi}\frac{A}{\epsilon^2}+
	\left(c_2 + c_3 \frac{A}{4\pi} \mu^2 + c_4 \frac{A}{4\pi} H^2\right)\ln\frac{1}{H \epsilon}
	+ c_5 \frac{A}{4\pi} H^2 + \frac{c_6}{2} \ln\left(A H^2\right) +\textrm{const}.
\end{equation}
The first line contains the UV divergent terms, while the second contains the finite ones. The Hubble constant $H$ in the arguments of the logarithms serves as a reference scale in order to make them dimensionless. In \cite{Maldacena:2012xp}, the emphasis was put on the coefficient $c_6$, for which the dependence on the mass $\mu$ was computed. Other terms were computed in \cite{Boutivas:2024sat,Boutivas:2024lts} by the current authors for a massless field in the Bunch-Davies vacuum, applying the approach of Srednicki \cite{Srednicki:1993im}. It was verified with high numerical accuracy that $c_2=c_6=-1/90$, so that the corresponding two terms combine into  $ \ln(R/\epsilon)$, as in flat space. It was also found that $c_4= 1/3$ with very high accuracy. The most remarkable finding, though, was the presence of an additional term 
\begin{equation}\label{eq:IR}
S_{\rm IR}=	c_{\rm IR} \frac{A}{4\pi} H^2 \ln(L H),
\end{equation}
with $L$ the  size of the {\it overall system}. It was shown, both numerically and analytically, that $c_{\rm IR}=1/3$, so that the terms involving $c_4$ and $c_{\rm IR}$ can be combined. It must be pointed out that the term \eqref{eq:IR} is always subleading to the dominant $1/\epsilon^2$-term within the range of validity of the analysis of \cite{Boutivas:2024sat,Boutivas:2024lts}. As a result, it does not affect the scaling properties of the entanglement entropy, which should be bounded by the number of degrees of freedom of the smaller subsystem \cite{Page:1993df}. However, this term indicates a dependence on the overall-system size, even when $L$ exceeds the Hubble radius $1/H$, as long as $LH$ is not exponentially large. The effect bears a strong similarity to the EPR paradox and is caused by the strong squeezing of the wave function of the field modes as they are stretched beyond the horizon by the expansion.

In this work, we turn to another curved space of high interest, namely AdS space. Srednicki's seminal work \cite{Srednicki:1993im} established a profound connection between entanglement entropy and geometric boundaries in quantum field theory, revealing that the entanglement entropy of a free massless scalar field in flat space exhibits an area law. Extending this analysis to an AdS background is a natural, yet highly non-trivial, step, offering new insights on the intersection of quantum information, gravity, and holography. Unlike flat space, AdS features curvature, a natural length scale, and a boundary at infinity, both of which significantly alter the structure of correlations. Understanding entanglement entropy in this setting sheds light on how curvature affects quantum correlations. Furthermore, studying a free scalar field in AdS provides a tractable model to explore how the entanglement area law generalizes in curved space and how subleading corrections encode universal quantum properties. This analysis is also particularly interesting in comparison with dS space, given that results for the two spaces are related by analytical continuation, despite the key differences in the causal structure.

In the following, we study the entanglement entropy of a scalar field in anti-de Sitter (AdS) space, adopting global coordinates. We consider spherical regions around the origin of AdS and compute the expression equivalent to \eqref{eq:SEE_expansion_mald}. We note that our analysis is not related to the Ryu-Takayanagi conjecture \cite{Ryu:2006bv,Ryu:2006ef} in the context of the AdS/CFT correspondence, for which the entangling surface is located on the AdS boundary.\footnote{The holographic entanglement entropy is given by the Ryu-Takayanagi conjecture as the area of a minimal surface in the bulk that is anchored on the entangling surface at the boundary. In order to connect our results to this conjecture, we would have to select an entangling surface in AdS that would coincide with this minimal surface (see e.g. \cite{Sugishita:2016iel}).} We follow  the approach of Srednicki \cite{Srednicki:1993im} in order to compute numerically the coefficients of the UV divergent terms, parameterized as
\begin{equation}\label{eq:adsentropy}
	S_{\textrm{AdS}}=\frac{d_1}{4\pi}\frac{A}{\epsilon^2}+
	\left(d_2 + d_3 \frac{A}{4\pi} \mu^2 + d_4 \frac{A}{4\pi a^2} \right)\ln\frac{a}{\epsilon} + \textrm{finite},
\end{equation}
where $a$ is the AdS length. The coefficients $d_2$, $d_3$, and $d_4$ are universal, while $d_1$ depends on the choice of regulator. Since the divergent term proportional to $d_1$ is dominant, extracting precise information about the finite terms becomes particularly challenging. Moreover, the form of the finite terms themselves depends on the choice of regulator, and some choices may lead to a more complicated structure for these terms than others, making it difficult to perform specific numerical fits to determine them. For example, a drastic redefinition of $\epsilon$ in \eqref{eq:SEE_expansion_mald} would affect the finite term proportional to $c_5$. This would then hinder the extraction of information on the coefficient $c_6$ through numerical fits. The regulator we use in the following is particularly sensitive to such effects. On the other hand, it provides a very efficient way of computing the divergent terms with high precision. 

The structure of the paper is as follows: In section \ref{sec:EE_in_AdS}, we set up the calculation of the entanglement entropy. Section \ref{sec:results} presents the numerical results, followed by a discussion in section \ref{sec:discussion}.

\section{Entanglement Entropy in AdS}
\label{sec:EE_in_AdS}
\subsection{General setup}
\label{subsec:Generalities}
We consider a scalar field on a fixed AdS background. We employ global coordinates, in which the metric of AdS$_{d+1}$ reads
\begin{equation}\label{eq:globalr}
	ds^2=-f(r)dt^2+\frac{1}{f(r)}dr^2+r^2d\Omega^2_{d-1},\qquad f(r)=1+\frac{r^2}{a^2},
\end{equation}
where $a$ is the AdS length. We are interested in the calculation of the entanglement entropy of a free scalar field at its ground state, when a spherical region around the origin of AdS is traced out at a given time. We implement the methodology of \cite{Srednicki:1993im}. We expand the field in spherical harmonics and discretize the radial coordinate in order to express the theory as a quantum mechanical system. Then, the entanglement entropy can be calculated using elementary quantum mechanics. The continuum limit is taken by increasing the number of lattice points in the discretized radial direction.

The action of a free scalar field $\phi(t,\vec{r})$ in AdS$_{d+1}$ is 
\begin{equation}\label{eq:action_initial}
	\mathcal{S}=\frac{1}{2}\int dt dr d\Omega_{d-1}r^{d-1}\left[\frac{1}{f(r)}\dot{\phi}^2-f(r)\left(\partial_r\phi\right)^2+\frac{1}{r^2}\phi\Delta_{d-1}\phi-\mu^2\phi^2\right],
\end{equation}
where a dot denotes a derivative with respect to time, $\Delta_{d-1}$ is the Laplacian on the unit $(d-1)$-dimensional sphere, and $\mu$ is the mass of the field. The conformal compactification of global AdS is obtained by introducing the tortoise coordinate $w$ according to
\begin{equation}\label{eq:tortoise_def}
	r=a\tan\frac{w}{a}\,,
\end{equation}
so that $w=0$ corresponds to the center of AdS and $w=a\pi/2$ to the conformal boundary. The metric now reads
\begin{equation}\label{eq:globalw}
	ds^2=\frac{1}{\cos^2\frac{w}{a}}\left(-dt^2+dw^2+a^2\sin^2\frac{w}{a} d\Omega^2_{d-1}\right)
\end{equation}
and the action of the scalar field becomes
\begin{equation}\label{eq:ScalarAction}
	\mathcal{S}=\frac{1}{2}\int dt \int_0^{a\frac{\pi}{2}} dw\int_{\textrm{S}^{d-1}} \hspace{-0.5cm} d\Omega_{d-1}\left(a\tan\frac{w}{a}\right)^{d-1}\left[\dot{\phi}^2-\left(\partial_w\phi\right)^2+\frac{\phi\Delta_{d-1}\phi}{a^2\sin^2\frac{w}{a}}-\frac{\mu^2\phi^2}{\cos^2\frac{w}{a}}\right].
\end{equation}

In order to trace out a spherical region around the origin of AdS, it is advantageous to expand the field using real hyper-spherical harmonics as
\begin{equation}\label{eq:mode_expansion}
	\phi\left(t,w,\hat{r}\right)=\frac{1}{a^{\frac{d-1}{2}}\tan^{\frac{d-1}{2}}\frac{w}{a}}\sum_{\ell,\vec{m}}\phi_{\ell\vec{m}}\left(t,w\right)Y_{\ell \vec{m}}\left(\hat{r}\right),
\end{equation}
where $\vec{m}=(m_1,\dots,m_{d-2})$ and $\hat{r}$ is a unit vector.\footnote{The real hyper-spherical harmonics $Y_{\ell \vec{m}}\left(\hat{r}\right)$ obey
\begin{equation}
		\Delta_{d-1}Y_{\ell \vec{m}}\left(\hat{r}\right)=-\ell\left(\ell+d-2\right)Y_{\ell \vec{m}}\left(\hat{r}\right)
\end{equation}
and the orthogonality condition
\begin{equation}
		\int_{\textrm{S}^{d-1}}\hspace{-0.5cm}d\Omega_{d-1}\,Y_{\ell \vec{m}}\left(\hat{r}\right)Y_{\ell^\prime \vec{m}^\prime}\left(\hat{r}\right)=\delta_{\ell,\ell^\prime}\delta_{\vec{m},\vec{m}^\prime}.
\end{equation}}
The original theory is divided into sectors that do not interact with each other. This is a consequence of the fact that the action assumes the form 
\begin{equation}
	\mathcal{S}=\sum_{\ell , \vec{m}}\mathcal{S}_{\ell , \vec{m}},
\end{equation}
where the action of each $(\ell,\vec{m})$-sector reads
\begin{multline}\label{eq:Action_lm}
	\mathcal{S}_{\ell\vec{m}}=\frac{1}{2}\int dt \int_0^{a\frac{\pi}{2}} dw  \left[\dot{\phi}_{\ell\vec{m}}^2-\left(\partial_w\phi_{\ell\vec{m}}\right)^2-\frac{1}{a^2}\left(\frac{\nu^2-\frac{1}{4}}{\sin^2\frac{w}{a}}+\frac{\kappa^2-\frac{1}{4}}{\cos^2\frac{w}{a}}\right)\phi^2_{\ell\vec{m}}\right]\\
	+\frac{d-1}{2}\int dt \int_0^{a\frac{\pi}{2}} dw\, \partial_w\left(\frac{\phi^2_{\ell\vec{m}}}{a\sin\frac{2w}{a}}\right),
\end{multline}
with the parameters $\nu$ and $\kappa$ given by
\begin{equation}\label{eq:nu_kappa_def}
	\nu=\ell+\frac{d}{2}-1\,,\qquad \kappa=\sqrt{\mu^2 a^2+\frac{d^2}{4}}\,.
\end{equation}
The action of each sector \eqref{eq:Action_lm} describes a field in $1+1$ dimensions with a trigonometric P\"oschl-Teller potential. The fields $\phi_{\ell\vec{m}}$ are canonically normalized, as we have included an appropriate factor in the expansion \eqref{eq:mode_expansion}.

To specify the physical modes, we also need to impose conditions for the behavior of $\phi_{\ell\vec{m}}$ at the endpoints $0$ and $a\pi/2$. These are related respectively to a regularity condition at $0$ and a boundary condition on the conformal boundary at $a\pi/2$, both of which must be imposed on the scalar field $\phi$ rather than on the modes $\phi_{\ell\vec{m}}$ -- imposing conditions directly on $\phi_{\ell\vec{m}}$ does not ensure general covariance. The invariant regularity condition at the origin of AdS is that the scalar field $\phi$ remains finite there. Through the expansion \eqref{eq:mode_expansion}, this translates to the condition
\begin{equation}\label{eq:regularity}
	\phi_{\ell\vec{m}}=\mathcal{O}(w^{\frac{d-1}{2}})\,,\qquad w\to0^+\,.
\end{equation}  
It follows that the surface term at $w=0$ in \eqref{eq:Action_lm} vanishes for regular solutions and $d\geq 3$.

The permissible boundary conditions at $w=a\pi/2$ depend on the value of the mass parameter $\mu^2$. For $\kappa>0$, the equations of motion following from the action \eqref{eq:ScalarAction} imply that, near the conformal boundary at $w=a\pi/2$, the most general asymptotic behavior of the scalar field $\phi$ takes the form
\begin{equation}\label{eq:AdSasymptotics}
	\phi\left(t,w,\hat{r}\right)\sim \phi_-(t,\hat{r})\left(a\frac{\pi}{2}-w\right)^{d-\Delta}+\cdots+\phi_+(t,\hat{r})\left(a\frac{\pi}{2}-w\right)^{\Delta}+\cdots\,,
\end{equation}
where $\phi_\pm(t,\hat{r})$ are arbitrary functions and $\Delta=\kappa+d/2$. For $\kappa=0$, in which case the Breitenlohner-Freedman (BF) bound  \cite{Breitenlohner:1982bm,Breitenlohner:1982jf} is saturated, the near boundary behavior of the scalar field is instead 
\begin{equation}\label{eq:AdSasymptoticsBF}
	\phi\left(t,w,\hat{r}\right)\sim \left(a\frac{\pi}{2}-w\right)^{d/2}\left(\phi_-(t,\hat{r})\log\left(a\frac{\pi}{2}-w\right)+\cdots+\phi_+(t,\hat{r})\right)+\cdots\,.
\end{equation}

These expansions determine the asymptotic behavior of the modes $\phi_{\ell\vec{m}}(t,w)$ defined through \eqref{eq:mode_expansion}. In particular, the asymptotic form of $\phi_{\ell\vec{m}}(t,w)$ is given by
\begin{equation}
	\phi_{\ell\vec{m}}(t,w)\sim \left(a\frac{\pi}{2}-w\right)^{\frac{1-d}{2}}a^{\frac{d-1}{2}}\int_{\textrm{S}^{d-1}}\hspace{-0.5cm}d\Omega_{d-1}\,Y_{\ell \vec{m}}\left(\hat{r}\right)\phi\left(t,w,\hat{r}\right)\,.
\end{equation}
It follows that for $\kappa>0$, 
\begin{equation}\label{eq:near_boundary_exp_1} 
	\phi_{\ell\vec{m}}(t,w)\sim \phi^-_{\ell\vec{m}}(t)\left(a\frac{\pi}{2}-w\right)^{\frac{1}{2}-\kappa}+\cdots+\phi^+_{\ell\vec{m}}(t)\left(a\frac{\pi}{2}-w\right)^{\frac{1}{2}+\kappa}+\cdots\,,
\end{equation}
where
\begin{equation}\label{eq:ModeDecomposition}
	\phi^\pm_{\ell\vec{m}}(t)\equiv a^{\frac{d-1}{2}}\int_{\textrm{S}^{d-1}}\hspace{-0.5cm}d\Omega_{d-1}\,Y_{\ell \vec{m}}\left(\hat{r}\right)\phi_\pm\left(t,\hat{r}\right)\,.
\end{equation}
Similarly, for $\kappa=0$,
\begin{equation}\label{eq:near_boundary_exp_2} 
	\phi_{\ell\vec{m}}(t,w)\sim \left(a\frac{\pi}{2}-w\right)^{\frac{1}{2}}\left(\phi^-_{\ell\vec{m}}(t)\log\left(a\frac{\pi}{2}-w\right)+\cdots+\phi^+_{\ell\vec{m}}(t)+\cdots\right)\,.
\end{equation}

The boundary conditions that can be consistently imposed on the field $\phi$ follow from the infinitesimal variation of the action \eqref{eq:ScalarAction}, which takes the form
\begin{multline}\label{eq:ScalarActionvar}
	\delta\mathcal{S}=\int dt \int_0^{a\frac{\pi}{2}} dw\int_{\textrm{S}^{d-1}} \hspace{-0.5cm} d\Omega_{d-1}\left(a\tan\frac{w}{a}\right)^{d-1}\bigg\{-\ddot{\phi}+\left(a\tan\frac{w}{a}\right)^{1-d}\partial_w\left[\left(a\tan\frac{w}{a}\right)^{d-1}\partial_w\phi\right]\\
	+\frac{\Delta_{d-1}\phi}{a^2\sin^2\frac{w}{a}}-\frac{\mu^2\phi}{\cos^2\frac{w}{a}}\bigg\}\delta\phi+\int_0^{a\frac{\pi}{2}} dw\int_{\textrm{S}^{d-1}} \hspace{-0.5cm} d\Omega_{d-1}\left(a\tan\frac{w}{a}\right)^{d-1}\dot\phi\,\delta\phi\bigg|_{t=-\infty}^{t=+\infty}\\
	-\lim_{w\to a\frac{\pi}{2}}\int dt \int_{\textrm{S}^{d-1}} \hspace{-0.5cm} d\Omega_{d-1}\left(a\tan\frac{w}{a}\right)^{d-1}\partial_w\phi\,\delta\phi\,.
\end{multline}
Notice that the general asymptotic form \eqref{eq:AdSasymptotics} and \eqref{eq:AdSasymptoticsBF} of the solutions implies that the boundary term in the last line is generally divergent. A well-posed variational problem in AdS requires the addition of local and covariant boundary counterterms, $\mathcal{S}_{\rm ct}$, that can be determined algorithmically via holographic renormalization \cite{deHaro:2000vlm,Skenderis:2002wp,Papadimitriou:2004ap,Papadimitriou:2016yit}. 

The boundary counterterms $\mathcal{S}_{\rm ct}$ are a polynomial in conformal powers of the Laplacian on the boundary and their explicit form depends on the dimension $d$ and the AdS mass parameter $\mu^2$ \cite{deHaro:2000vlm,Skenderis:2002wp,Papadimitriou:2004ap,Papadimitriou:2016yit}. Their effect is that the boundary term in the variation of the total action $\mathcal{S}+\mathcal{S}_{\rm ct}$ admits a finite limit as $w\to a\pi/2$, namely, for $\kappa>0$,\footnote{In general, the coefficient of $\delta\phi_-$ is $(d-2\Delta)\phi_+/a+C[\phi_-]$, where $C[\phi_-]$ is a dimension and mass dependent linear function of $\phi_-$ \cite{deHaro:2000vlm,Skenderis:2002wp,Papadimitriou:2004ap,Papadimitriou:2016yit}. However, the form of this function does not affect the present analysis.}
\begin{equation}\label{eq:TotalActionVariation}
	\delta(\mathcal{S}+\mathcal{S}_{\rm ct})\sim\frac{(d-2\Delta)}{a}\int dt \int_{\textrm{S}^{d-1}} \hspace{-0.5cm} d\Omega_{d-1}\phi_+\,\delta\phi_-\,,
\end{equation}
where $\phi_\pm(t,\hat r)$ are the arbitrary coefficients parameterizing the near boundary expansion \eqref{eq:AdSasymptotics}. An analogous expression holds for $\kappa=0$. 

The boundary term \eqref{eq:TotalActionVariation} vanishes provided we impose the Dirichlet boundary condition $\delta\phi_-=0$, which, via \eqref{eq:ModeDecomposition}, translates to $\delta\phi^-_{\ell\vec{m}}(t)=0$. Hence, with Dirichlet boundary conditions, the dynamical modes are proportional to $\phi_+$ and are normalizable for all values of the mass parameter $\mu^2$. Since we are focusing on a free scalar field here, without loss of generality, we may set the background value of $\phi_-(t,\hat r)$ or $\phi^-_{\ell\vec{m}}(t)$ to zero for the computation of the entanglement entropy.   

Alternative boundary conditions are possible for a certain range of the mass parameter $\mu^2$. In particular, if the mass parameter is in the window
\begin{equation}
	-\frac{d^2}{4}\leq \mu^2a^2< -\frac{d^2}{4}+1\,,\qquad\left(0\leq\kappa< 1\,,\quad \frac{d}{2}\leq\Delta<\frac{d}{2}+1\right)\,,
\end{equation}
then both modes $\phi_\pm(t,\hat r)$ are normalizable and alternative boundary conditions exist \cite{Klebanov:1999tb}. In particular, adding a further finite boundary term to the action, we may impose Neumann, $\delta\phi_+=0$, or mixed, $\delta((2\Delta-d)\phi_+/a-2\alpha\, \phi_-)=0$, boundary conditions \cite{Papadimitriou:2007sj}.

It should be stressed that, due to the conformal nature of the AdS boundary, imposing a condition on the field itself is not necessarily equivalent to imposing a consistent boundary condition, or indeed any boundary condition at all. For example, setting $\phi=0$ at $w=a\pi/2$ imposes Dirichlet boundary conditions only when $\mu^2a^2\geq 0$, or $\Delta\geq d$. Similarly, requiring that $\phi_{\ell\vec{m}}=0$ at $w=a\pi/2$ imposes Dirichlet boundary conditions only when $\mu^2 a^2\geq -\frac{d^2-1}{4}$, or $\Delta\geq \frac{d+1}{2}$, or $\kappa\geq 1/2$. This observation will be particularly important for the subsequent analysis, since the discretized eigenvalue problem will be solved imposing conditions only on the modes $\phi_{\ell\vec{m}}$, rather than on the asymptotic coefficients $\phi^\pm_{\ell\vec{m}}$ at $w=a\pi/2$. Notice that the bound for the condition $\phi_{\ell\vec{m}}=0$ at $w=a\pi/2$ to impose a boundary condition is saturated for a conformally coupled scalar with mass parameter $\mu^2 a^2=-(d/2)^2+1/4$, which translates to $\Delta=\frac{d+1}{2}$ and $\kappa=1/2$.

Once suitable boundary conditions have been imposed, the variational problem leads to the following second-order equation for the modes $\phi_{\ell\vec{m}}$  
\begin{equation}
	\ddot{\phi}_{\ell\vec{m}}-\partial^2_w\phi_{\ell\vec{m}}+\frac{1}{a^2}\left(\frac{\nu^2-\frac{1}{4}}{\sin^2\frac{w}{a}}+\frac{\kappa^2-\frac{1}{4}}{\cos^2\frac{w}{a}}\right)\phi_{\ell\vec{m}}=0\,.
\end{equation}
Writing $\phi_{\ell\vec{m}}(t,w)=e^{\pm iEt} \varphi_{\ell\vec{m}}(w)$ leads to an effective time-independent \Schrodinger equation with a trigonometric P\"oschl-Teller potential, namely
\begin{equation}\label{eq:eigensystem}
	-\partial^2_w\varphi_{\ell\vec{m}}+\frac{1}{a^2}\left(\frac{\nu^2-\frac{1}{4}}{\sin^2\frac{w}{a}}+\frac{\kappa^2-\frac{1}{4}}{\cos^2\frac{w}{a}}\right)\varphi_{\ell\vec{m}} = E^2 \varphi_{\ell\vec{m}} .
\end{equation}

The unique solution of this equation that satisfies the regularity condition \eqref{eq:regularity} at $w=0$ is given by 
\begin{multline}\label{eq:spectum_K}
	\varphi_{\ell\vec{m}}(w;s)=\frac{c_s}{\sqrt{a}}\frac{\Gamma(\nu+s+1)}{\Gamma(s+1)\Gamma(\nu+1)} \sin^{\nu+\frac{1}{2}}\left(\frac{w}{a}\right)\cos^{\kappa+\frac{1}{2}}\left(\frac{w}{a}\right)\\
	\times {}_2F_1\left(-s,s+\kappa+\nu+1,\nu+1,\sin^2\frac{w}{a}\right),
\end{multline}
where ${}_2F_1(\alpha,\beta,\gamma,z)$ is the Gauss hypergeometric function and $s$ is a spectral parameter that determines the energy via the relation
\begin{equation}\label{eq:eigenvalues}
	E_s = \frac{1}{a} \left(2s+\kappa+\nu+1\right)\,.
\end{equation}
Indeed, near $w=0$ the solution \eqref{eq:spectum_K} takes the form  
\begin{equation}\label{eq:asympt1b} 
	\varphi_{\ell\vec{m}}(w;s) = \frac{c_s}{\sqrt{a}}\frac{\Gamma\left(s+\nu+1\right)}{\Gamma\left(s+1\right)\Gamma\left(\nu+1\right)} \left(\frac{w}{a}\right)^{\nu+\frac{1}{2}} 	+{\mathcal{ O}}\left(  \left(\frac{w}{a}\right)^{\nu+\frac{5}{2}}\right)\,,
\end{equation}
which is compatible with the regularity condition \eqref{eq:regularity}, since $\nu+\frac12=\frac{d-1}{2}+\ell$ and $\ell\geq 0$. 

Let us next examine the near boundary behavior of the regular solution \eqref{eq:spectum_K}. Expanding this for $w^\prime= a\frac{\pi}{2}-w\ll a$ we obtain
\begin{multline}\label{eq:asympt1a}
	\varphi_{\ell\vec{m}}(w;s) =-\frac{c_s}{\sqrt{a}}\frac{1}{\sin(\pi\kappa)}\left[\frac{\pi\left(\frac{w^\prime}{a} \right)^{\kappa+\frac{1}{2}}}{\Gamma(s+1)\Gamma(-s-\kappa)\Gamma(\kappa+1)}+{\mathcal{ O}}\left(\left(\frac{w^\prime}{a}\right)^{\kappa+\frac{5}{2}}\right)\right. \\
\left.+\frac{\sin(s\pi)\Gamma(s+\nu+1)\left(\frac{w^\prime}{a} \right)^{-\kappa+\frac{1}{2}}}{\Gamma(1-\kappa)\Gamma(s+\kappa+\nu+1)}+{\mathcal{ O}}\left( \left(\frac{w^\prime}{a}\right)^{-\kappa+\frac{5}{2}}\right)\right]\,.
\end{multline}
This asymptotic form implies that the solution \eqref{eq:spectum_K} is normalizable for any $\kappa>0$ provided $s=n$ is a non-negative integer, or for any $s$ provided $0\leq\kappa< 1$ \cite{Ishibashi:2004wx}.

Comparing the expansion \eqref{eq:asympt1a} with the general asymptotic form \eqref{eq:near_boundary_exp_1}, we conclude that the case $s=n$ corresponds to Dirichlet boundary conditions, since $\phi^-_{\ell\vec{m}}(t)=0$. As expected, this boundary condition is admissible for any value of the mass parameter $\mu^2$ that satisfies the BF bound. In this case, the hypergeometric function in the solution \eqref{eq:spectum_K} reduces to a Jacobi polynomial, namely
\begin{equation}\label{eq:spectum_K_int}
	\varphi_{\ell\vec{m}}(w;n)=\frac{c_n}{\sqrt{a}} \sin^{\nu+\frac{1}{2}}\left(\frac{w}{a}\right)\cos^{\kappa+\frac{1}{2}}\left(\frac{w}{a}\right) P_{n}^{(\nu,\kappa)}\left(\cos\left(\frac{2w}{a}\right)\right)\,,
\end{equation}
and the solutions can be normalized to 1 by setting 
\begin{equation}
 	c_n^2= 2\left(2n+\kappa+\nu+1\right)\frac{\Gamma(n+\kappa+\nu+1)\Gamma(n+1)}{\Gamma(n+\kappa+1)\Gamma(n+\nu+1)}\,.
\end{equation}

\begin{figure}[t]
	\centering
	\begin{picture}(55,33.5)
	\put(0,0){\includegraphics[width=0.55\textwidth]{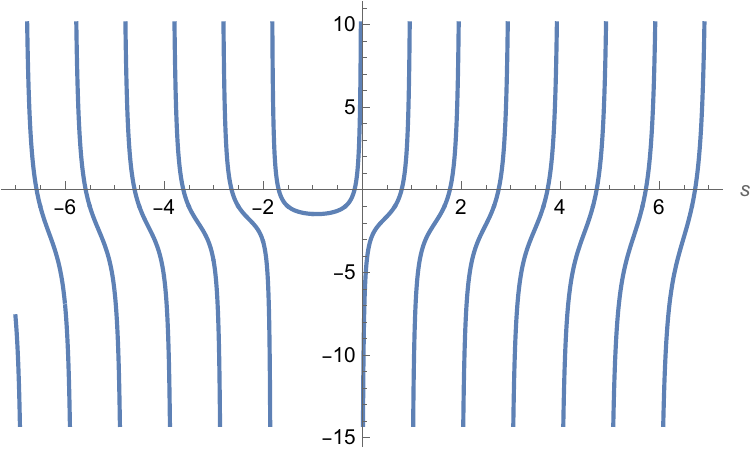} }
\end{picture}
\caption{Plot of the l.h.s. of \eqref{eq:mixed_spectrum} as a function of the spectral parameter $s$ for $\nu=1/2$, $\kappa=1/3$ and $A_\kappa=1$. The zeros of this function determine the spectrum $s_n$, $n\in \mathbb{Z}$, for mixed boundary conditions. For any value of $A_\kappa<+\infty$ that satisfies \eqref{eq:positivity}, both positive and negative roots must be included in the spectrum. For $A_\kappa\to+\infty$, corresponding to Dirichlet boundary conditions, the non-negative roots become integers, while the negative roots must be discarded.}
\label{fig:plot_mixed_spectrum}
\end{figure}

If $0\leq\kappa< 1$, the spectral parameter $s$ may take values other than a non-negative integer.\footnote{We assume $0<\kappa< 1$ in the discussion, since the case $\kappa=0$ requires special treatment.} In that case, both modes $\phi^\pm_{\ell\vec{m}}(t)$ are nonzero and, therefore, the regular solution \eqref{eq:spectum_K} satisfies mixed boundary conditions. For a free scalar, the relation between the modes can only be linear, and so mixed boundary conditions take the form \cite{Papadimitriou:2007sj}
\begin{equation}\label{eq:mixed_bc}
	\frac{2\Delta-d}{a}\phi_+-2\alpha\, \phi_-=0\,,	
\end{equation}	
where $\alpha$ is a fixed constant that defines the choice of mixed boundary condition. The case $\alpha=0$ corresponds to Neumann boundary conditions, while $\alpha\to\infty$ corresponds to Dirichlet. Reading off the modes $\phi^\pm_{\ell\vec{m}}(t)$ from the asymptotic solution \eqref{eq:asympt1a} and inserting them in the boundary condition \eqref{eq:mixed_bc} results in the following equation for the spectral parameter $s$:
\begin{equation}\label{eq:mixed_spectrum}
	\frac{\pi\,\Gamma(s+\kappa+\nu+1)}{\sin(s\pi)\Gamma(s+\nu+1)\Gamma(s+1)\Gamma(-s-\kappa)}-A_\kappa=0\,,\qquad A_\kappa\equiv\frac{\Gamma(\kappa)}{\Gamma(1-\kappa)}a^{2\kappa+1}\alpha\,,	
\end{equation}
where $A_\kappa$ is a fixed dimensionless parameter characterizing the mixed boundary condition. Given $\nu$, $\kappa$, and $A_\kappa$, this equation has a discrete set of solutions $s_n$, $n\in \mathbb{Z}$, corresponding to the spectrum for the specific boundary condition. An example is shown in Fig.~\ref{fig:plot_mixed_spectrum}.
  
As analyzed in detail in \cite{Ishibashi:2004wx,Barroso:2019cwp} (see also \cite{SAEbook}), the mixed boundary conditions \eqref{eq:mixed_bc} that are admissible for $\kappa$ in the range $0\leq\kappa< 1$ correspond to a one-parameter family of self-adjoint extensions of the \Schrodinger Hamiltonian \eqref{eq:eigensystem}. The self-adjoint extensions are parameterized by $A_\kappa$ in \eqref{eq:mixed_spectrum}. These extensions  result in a positive definite Hamiltonian provided the parameter $A_\kappa$ satisfies 
\begin{equation}\label{eq:positivity}
	A_\kappa\geq -\left|\frac{\Gamma(-\kappa)}{\Gamma(\kappa)}\right|\frac{\Gamma(1+\kappa)\Gamma\left(\frac{\nu+\kappa+1}{2}\right)^2}{\Gamma(1-\kappa)\Gamma\left(\frac{\nu-\kappa+1}{2}\right)^2}\,.
\end{equation}

Some remarks are in order at this point:
\begin{itemize}
	\item Requiring that $\phi_{\ell\vec{m}}=0$ at $w=a\pi/2$, as we will do in the numerical analysis below, imposes Dirichlet boundary conditions for $\kappa\geq 1/2$, while for $0\leq \kappa <1/2$ both Dirichlet and mixed boundary conditions are allowed. 
	\item 
	The action of an $(\ell,\vec{m})$-sector \eqref{eq:Action_lm} does not depend on $\vec{m}$, which implies that all modes with given $\ell$ contribute equally to the entanglement entropy. 
	\item 
	The modes $\varphi_{\ell\vec{m}}(w;n)$, $n\in \mathbb{Z}^+$, in the case of Dirichlet boundary conditions, and $\varphi_{\ell\vec{m}}(w;s_n)$, $n\in \mathbb{Z}$, in the case of mixed boundary conditions form a complete basis, on which the field can expanded.
	\item 
	 The BF stability bound \cite{Breitenlohner:1982bm,Breitenlohner:1982jf}
	\begin{equation}
		\mu^2 \geq -\frac{d^2}{4a^2}
	\end{equation}
is saturated for $\kappa=0$. Below the bound, the parameter $\kappa$ becomes imaginary. The eigenvalue problem \eqref{eq:eigensystem} has solutions whose eigenvalues are not bounded from below, thus signaling the presence of an instability \cite{Ishibashi:2004wx}.
	\item 
	A field with a non-minimal coupling $-\frac{1}{2}\xi R \phi^2$ in AdS develops an effective mass
	\begin{equation}
		\mu^2_{\rm eff}=\mu^2-\xi\frac{d(d+1)}{a^2}\,.
	\end{equation}
	For $\xi=\frac{d-1}{4d}$, which corresponds to a Weyl-invariant theory, we have $\kappa=\sqrt{\mu^2a^2+\frac{1}{4}}$ for any value of $d$. Thus, a conformally coupled, massless scalar corresponds to $\kappa=1/2$ and the term involving the cosine in \eqref{eq:Action_lm} is absent. This Lagrangian density coincides with that of a conformally coupled, massless scalar field on the sphere S$^d$, which indicates that the entropies in the two theories are closely related. However, it should be kept in mind that $0\leq w/a\leq \pi/2$ in AdS, whereas the analogous angle in S$^d$ takes values up to $\pi$.
\end{itemize}

\subsection{Discretization}
\label{subsec:Discretization}
In order to calculate the entanglement entropy, we employ the Hamiltonian corresponding to the action \eqref{eq:Action_lm}.\footnote{The proper definition of the Hamiltonian requires appropriate boundary terms according to the discussion in the previous subsection. However, for the Dirichlet boundary conditions that we consider in the following, these boundary terms do not affect the analysis.} The Hamiltonian reads
\begin{equation}
	H_{\ell\vec{m}}=\frac{1}{2}\int_0^{a\frac{\pi}{2}} dw \left[\pi^2_{\ell\vec{m}}+\left(\partial_w\phi_{\ell\vec{m}}\right)^2+\frac{1}{a^2}\left(\frac{\nu^2-\frac{1}{4}}{\sin^2\frac{w}{a}}+\frac{\kappa^2-\frac{1}{4}}{\cos^2\frac{w}{a}}\right)\phi^2_{\ell\vec{m}}\right],
\end{equation}
where $\pi_{\ell\vec{m}}=\dot{\phi}_{\ell\vec{m}}$. For $1/\mu\gg a\gg w$, the Hamiltonian assumes the same form as for a massless field in flat space. We expect then that the leading contributions to the entropy of a light field for entangling radii much smaller than the AdS length should coincide with those on a flat background. 

Our approach is based on the discretization of the theory, performed using the following scheme:
\begin{equation}\label{eq:discretization}
	w=\epsilon\, i,\qquad a=\frac{2}{\pi}\epsilon(N+1), \qquad \int_0^{a\frac{\pi}{2}} dw \rightarrow \epsilon\sum_{i=0}^{N+1},
\end{equation}
while the modes of the field are redefined as
\begin{equation}
	\phi_{\ell\vec{m}}(t,w)\rightarrow \frac{1}{\sqrt{\epsilon}}\phi_{\ell\vec{m},i}(t),\qquad \pi_{\ell\vec{m}}(t,w)\rightarrow \frac{1}{\sqrt{\epsilon}}\pi_{\ell\vec{m},i}(t).
\end{equation}
The Hamiltonian becomes
\begin{multline}
	H_{\ell\vec{m}}=\frac{1}{2}\sum_{i} \Bigg[\pi^2_{\ell\vec{m},i}+\frac{\left(\phi_{\ell\vec{m},i+1}-\phi_{\ell\vec{m},i}\right)^2}{\epsilon^2}\\
	+\left(\frac{\nu^2-\frac{1}{4}}{\sin^2\frac{\pi i}{2(N+1)}}+\frac{\kappa^2-\frac{1}{4}}{\cos^2\frac{\pi i}{2(N+1)}}\right)\frac{\pi^2\phi^2_{\ell\vec{m},i}}{4\epsilon^2(N+1)^2}\Bigg].
\end{multline}

We impose the endpoint conditions 
\begin{equation}
\phi_{\ell\vec{m},0}(t)=\phi_{\ell\vec{m},N+1}(t)=0,\qquad \pi_{\ell\vec{m},0}(t)=\pi_{\ell\vec{m},N+1}(t)=0.
\end{equation}
The conditions at $i=0$ enforce regularity at the origin of AdS$_4$. However, as we saw in the previous subsection, the conditions at $i=N+1$ impose a specific boundary condition (Dirichlet) only when $\kappa\geq 1/2$. For $0\leq \kappa <1/2$, these endpoint conditions are satisfied, not only by Dirichlet, but also by any boundary condition of the form \eqref{eq:mixed_bc}. This fact renders the discretization of the problem significantly more challenging when $0\leq \kappa <1/2$, since one must fix the asymptotic form of the eigenfunctions of the Hamiltonian, and not just their behavior at the endpoint. For this reason, we will limit our current analysis to the case $\kappa\geq 1/2$. Imposing the endpoint conditions results in a system of $N$ interacting harmonic oscillators, with Hamiltonian
\begin{equation}
	H_{\ell\vec{m}}=\frac{1}{2}\sum_{i=1}^N \pi^2_{\ell\vec{m},i}
	+\frac{1}{2}\sum_{i,j=1}^N \phi_{\ell\vec{m},i}\, K_{ij} \, \phi_{\ell\vec{m},j} \, .
\end{equation}
The interactions are determined by the coupling matrix
\begin{equation}
	K_{ij}=\frac{1}{\epsilon^2}\left[\left(2+\left(\frac{\nu^2-\frac{1}{4}}{\sin^2\frac{\pi i}{2(N+1)}}+\frac{\kappa^2-\frac{1}{4}}{\cos^2\frac{\pi i}{2(N+1)}}\right)\frac{\pi^2}{4(N+1)^2}\right)\delta_{i,j}-\delta_{i+1,j}-\delta_{i,j+1}\right],
\end{equation}
where we have accounted for the endpoint conditions at the ends of the chain. The above coupling matrix is convenient for numerical calculations in which dimensionful quantities are expressed in terms of the lattice spacing $\epsilon$, which effectively is set equal to 1. For our problem, it is more convenient to use the AdS length $a$ as a reference scale. This can be achieved by writing the coupling matrix as
\begin{equation}\label{eq:coupling matrix}
	K_{ij}=\frac{1}{a^2}\left[\frac{4(N+1)^2}{\pi^2}\left(2\delta_{i,j}-\delta_{i+1,j}-\delta_{i,j+1}\right)+\left(\frac{\nu^2-\frac{1}{4}}{\sin^2\frac{\pi i}{2(N+1)}}+\frac{\kappa^2-\frac{1}{4}}{\cos^2\frac{\pi i}{2(N+1)}}\right)\delta_{i,j}\right]
\end{equation}
and setting $a=1$. The continuum limit is approached for $N\to \infty$. 

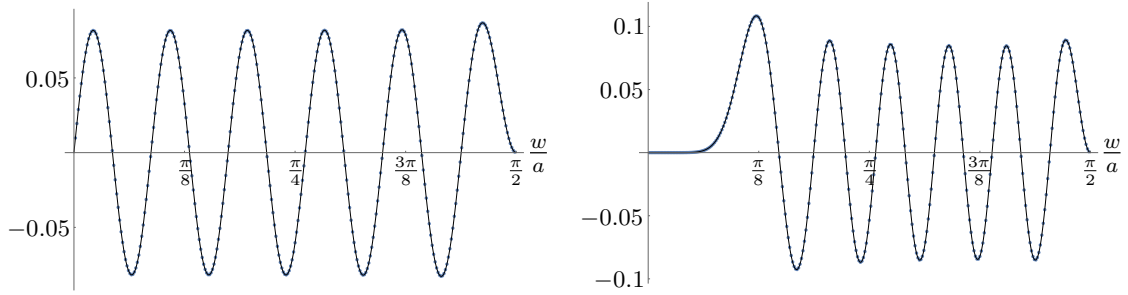
\begin{figure}[t]
	\centering
	\begin{picture}(97.5,25.5)
		\put(5,0){\includegraphics[width=0.4\textwidth]{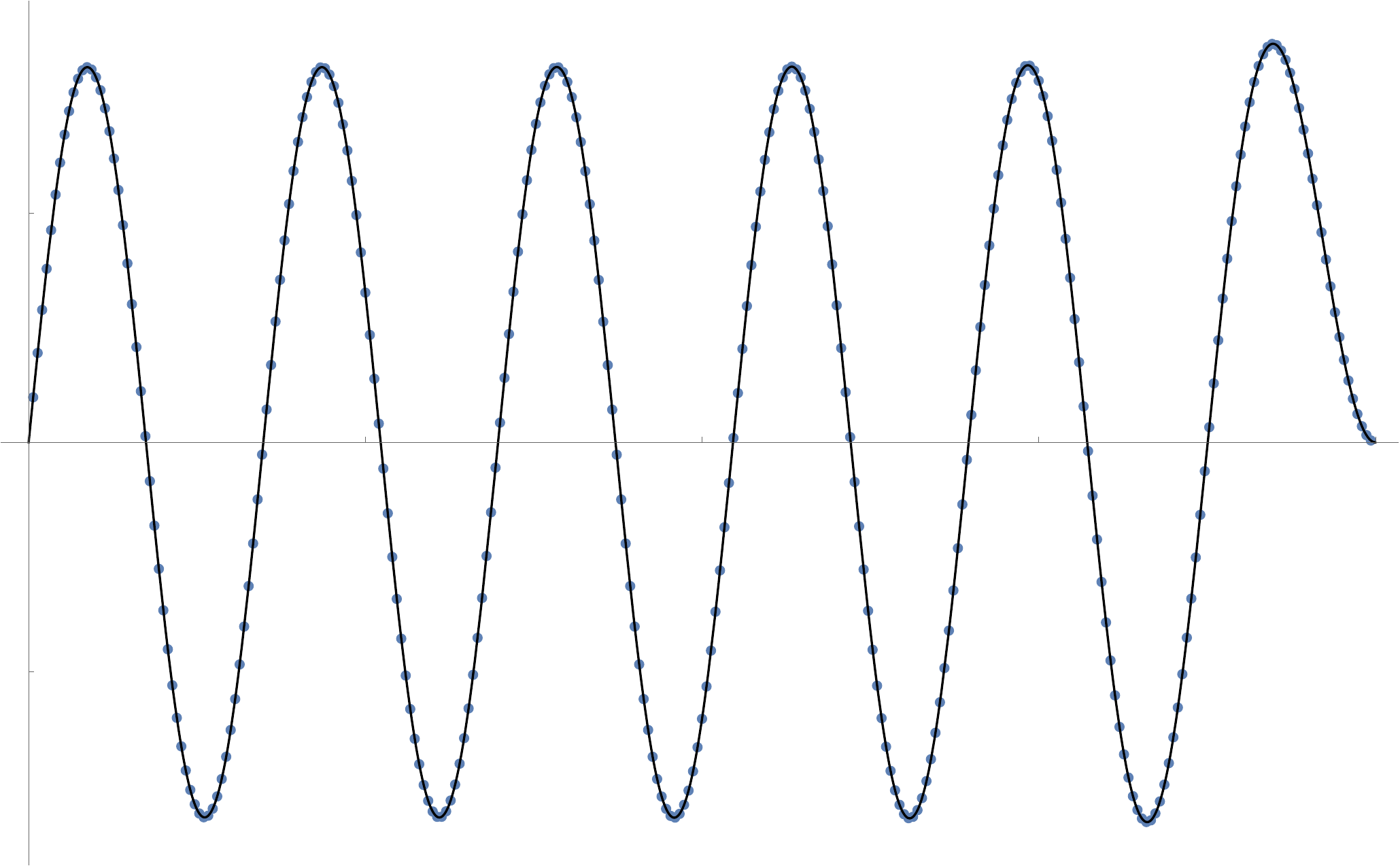} }
		\put(55,0.6){\includegraphics[width=0.4\textwidth]{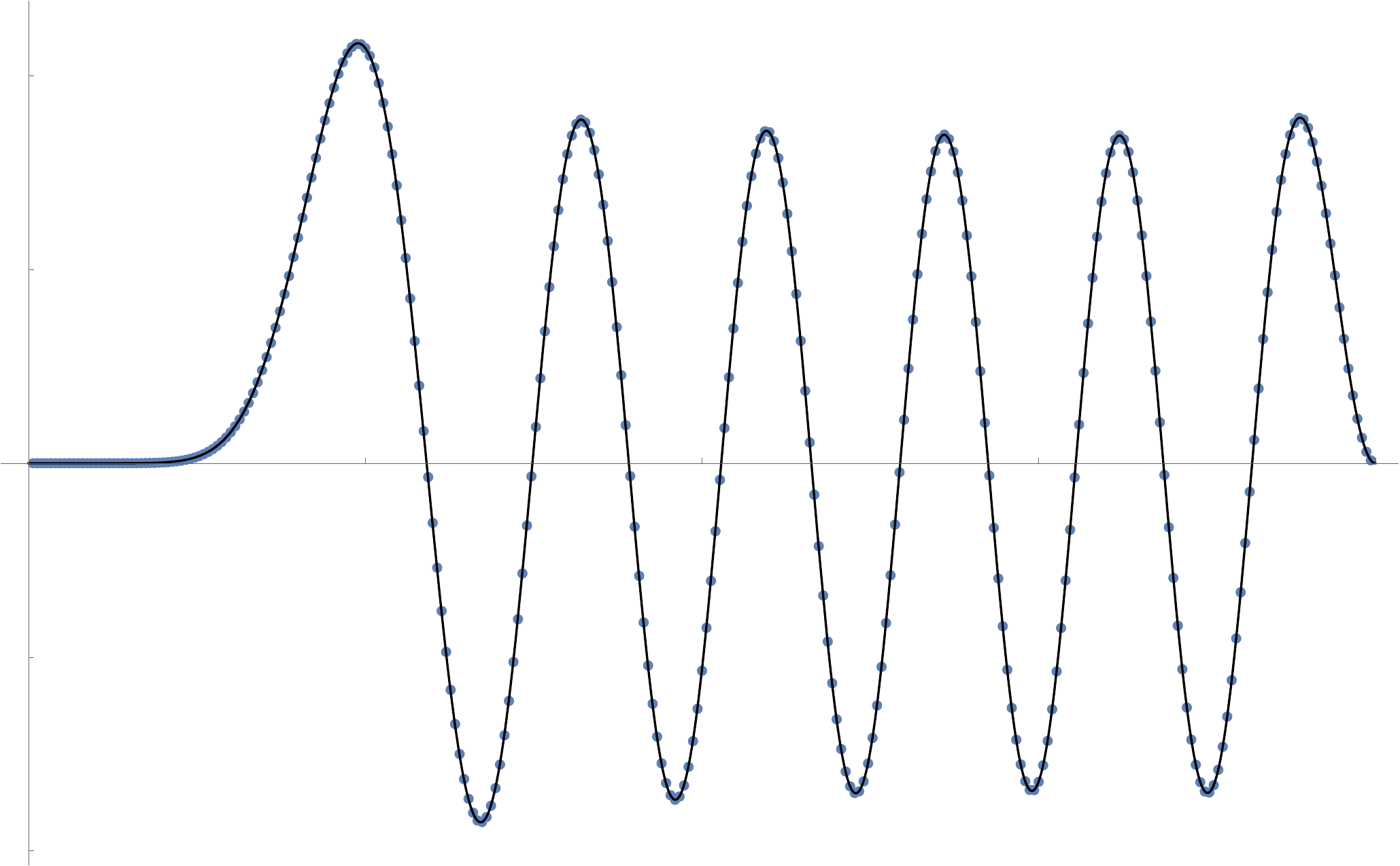} }
		\put(45.25,11.4){{\large $\frac{w}{a}$}}
		\put(95.2,11.4){{\large $\frac{w}{a}$}}
		\put(14.7,10){{\footnotesize $\frac{\pi}{8}$}}
		\put(24.3,10){{\footnotesize $\frac{\pi}{4}$}}
		\put(33.5,10){{\footnotesize $\frac{3\pi}{8}$}}
		\put(43.5,10){{\footnotesize $\frac{\pi}{2}$}}
		\put(64.6,10){{\footnotesize $\frac{\pi}{8}$}}
		\put(74.3,10){{\footnotesize $\frac{\pi}{4}$}}
		\put(83.4,10){{\footnotesize $\frac{3\pi}{8}$}}
		\put(93.5,10){{\footnotesize $\frac{\pi}{2}$}}
		\put(0,4.8){{\footnotesize $-0.05$}}
		\put(1.6,17.9){{\footnotesize $0.05$}}
		\put(51.1,0.3){{\footnotesize $-0.1$}}
		\put(49.9,5.8){{\footnotesize $-0.05$}}
		\put(51.6,16.9){{\footnotesize $0.05$}}
		\put(52.8,22.5){{\footnotesize $0.1$}}
	\end{picture}
	\caption{The eigenfunctions of the coupling matrix \eqref{eq:coupling matrix} with $N=299$ and the corresponding analytical modes of the continuous theory \eqref{eq:spectum_K} (continuous line) for $\ell=0$, $n=10$ (left plot) and $\ell=10$, $n=10$ (right plot) in the massless theory ($\mu=0$, $\kappa=3/2$).}
	\label{fig:plot1}
\end{figure}

For large $N$, the eigenfunctions of the matrix \eqref{eq:coupling matrix} should coincide with the modes \eqref{eq:spectum_K}. In Fig.~\ref{fig:plot1}, we display the eigenfunctions of the coupling matrix with $N=299$ and the corresponding analytical modes for $\ell=0$, $n=10$ (left plot) and $\ell=10$, $n=10$ (right plot) in the massless theory ($\mu=0$, $\kappa=3/2$). It is evident that there is an excellent agreement between the analytic formulae and the numerical results.

\subsection{Calculation of the Entropy}
The calculation proceeds similarly to Srednicki's analysis for a scalar field in flat space \cite{Srednicki:1993im}, with the only difference being the use of the modified coupling matrix $K$, adapted to the curved background and given by \eqref{eq:coupling matrix}. The ground state wave function of the discretized system for each $(\ell,\vec{m})$-sector is
\begin{equation}
	\Psi({{\bm \phi}_{\ell\vec{m}}})=\left(\det\frac{\Omega}{\pi}\right)^{1/4} e^{-\frac{1}{2} {\bm \phi}_{\ell\vec{m}}^T \Omega\, {\bm \phi}_{\ell\vec{m}}},
\end{equation}
where $\Omega$ is the positive square root of the matrix $K$ and ${\bm \phi}_{\ell\vec{m}}$ the column vector composed of the $\phi_{\ell\vec{m},j}$. Considering the oscillators $1$ to $n$ as subsystem $A$, and $n+1$ to $N$ as the complementary subsystem $C$, the entanglement entropy is
\begin{equation}
	S_{\ell\vec{m}}^{\textrm{EE}}=\sum_{i=1}^{N-n}\left(\frac{\sqrt{\lambda_i}+1}{2}\ln\frac{\sqrt{\lambda_i}+1}{2}-\frac{\sqrt{\lambda_i}-1}{2}\ln\frac{\sqrt{\lambda_i}-1}{2}\right),
\end{equation}
where $\lambda_i$ are the eigenvalues of the matrix
\begin{equation}
	\mathcal{M}=\left(\Omega^{-1}\right)_C\left(\Omega\right)_C.
\end{equation}
The matrix $\left(\Omega\right)_C$ is the $(N-n)\times(N-n)$ bottom-right block of $\Omega$ and similarly $\left(\Omega^{-1}\right)_C$ is the $(N-n)\times(N-n)$ bottom-right block of $\Omega^{-1}$. This result can be obtained by implementing the method of correlation functions for the calculation of the entanglement entropy \cite{Peschel:2002yqj}, which is equivalent to the approach of \cite{Srednicki:1993im}. (See \cite{Katsinis:2024gef,Katsinis:2023hqn} for additional information on this equivalence.) The overall entanglement entropy is obtained by summing the contributions of all the $(\ell,\vec{m})$-sectors, taking into account the degeneracy of these contributions. For AdS$_4$, i.e., for $d=3$, it follows that
\begin{equation}\label{eq:SEE_total}
	S_{\textrm{EE}}=\sum_{\ell=0}^\infty(2\ell+1)S_{\ell}^{\textrm{EE}}.
\end{equation}

Before describing the numerical calculation and presenting the results, there is a final issue that must be addressed. The index $n$ corresponds to the location of the last degree of freedom that belongs to system $A$, whereas $n+1$  corresponds to the location of the first degree of freedom that belongs to system $C$. The entangling surface lies between these two locations. The entangling surface in the continuous theory is determined by the relation $w = w_R$, where\footnote{This is in analogy with the definition $R\rightarrow\left(n+1/2\right)\epsilon$ in the flat-space theory \cite{Srednicki:1993im}.}
\begin{equation}
	\frac{w_R}{a}=\frac{\pi}{2}\frac{n+\frac{1}{2}}{N+1}.
\label{eq:wr}\end{equation}
Since we work with a fixed AdS scale, instead of a fixed lattice spacing $\epsilon$, it is advantageous for the numerical analysis to incorporate this shift by $1/2$ in the coupling matrix. Thus, we modify slightly the coupling matrix \eqref{eq:coupling matrix} and express it as
\begin{equation}\label{eq:coupling matrix_final}
	K_{ij}=\frac{1}{a^2}\left[\frac{4(N+1)^2}{\pi^2}\left(2\delta_{i,j}-\delta_{i+1,j}-\delta_{i,j+1}\right)+\left(\frac{\nu^2-\frac{1}{4}}{\sin^2\frac{\pi \left(i+\frac{1}{2}\right)}{2(N+1)}}+\frac{\kappa^2-\frac{1}{4}}{\cos^2\frac{\pi  \left(i+\frac{1}{2}\right)}{2(N+1)}}\right)\delta_{i,j}\right].
\end{equation}
Our numerical analysis verifies that this form of the matrix leads to a more efficient handling of the UV divergent terms and a faster convergence towards the continuum limit. However, the numerical analysis may also be performed without this shift, leading to the same results in the regime $\kappa\geq 1/2$ that we consider. In the regime $0\leq \kappa <1/2$, the shift significantly affects the form of the eigenfunctions because of their asymptotic behavior, and either it should be avoided or appropriate boundary conditions that take it into account should be applied.

\section{Numerical Analysis and Results}
\label{sec:numresults}
\subsection{Methodology}
We perform the numerical calculation for AdS$_4$ (i.e., $d=3$). We study the dependence of the entanglement entropy on the mass $\mu$ of the field, in contrast to \cite{Srednicki:1993im}, where only the massless case was considered. There is also a difference in how the continuum limit is approached.

In \cite{Srednicki:1993im}, the lattice spacing was kept fixed since the goal was to obtain results for the infinite-size continuous theory. This limit corresponds to $\epsilon\to 0$, $n\to\infty$ and $N\to\infty$ so that $R = n\epsilon$ is fixed, while $N\epsilon=L\to \infty$. Numerically, this limit is obtained as follows: For some specific values of $n$, e.g., the lattice points $15$ to $45$, one studies the entanglement entropy as a function of $N$. Through extrapolation, one obtains the values that correspond to infinite $N$. Since $N$ defines the dimensionless ratio $L/\epsilon$, this process corresponds to sending the size of the system to infinity. At the same time, the entanglement entropy becomes a function of only one parameter, namely $n$, which is the only dimensionless ratio left. In the continuous theory, the expression for the entanglement entropy matches that of the discretized theory when $n$ replaced by $R/\epsilon$. Thus, in order to isolate only the significant terms in the continuum limit, one fits the entanglement entropy with a series in $n$ and keeps only the non-negative powers of this series. The exact procedure is described in more detail in \cite{Lohmayer:2009sq}.

Here, we employ a different procedure: We use coordinates such that the total system corresponds to the finite interval between $0$ and $a\frac{\pi}{2}$. As a result, for our calculation, it is more natural to keep the AdS scale fixed, instead of the lattice spacing. In this way, we essentially bind the value of the UV cutoff and the number of the lattice points of the overall system through \eqref{eq:discretization}. The continuum limit is approached for $N\to \infty$.

Our choice of coordinates has a strong effect on the power-law divergent terms of the entanglement entropy. The AdS metric in global coordinates can be written either in the form \eqref{eq:globalr} or in the form \eqref{eq:globalw}, with the respective radial coordinates $r$ and $w$ related through \eqref{eq:tortoise_def}. The discretized version of the theory can be constructed through a radial lattice with constant spacing in terms of either coordinate. The two choices differ with respect to the number of degrees of freedom within a given proper length in the radial direction. This is reflected in the density of the entangled degrees of freedom, which affects the entropy. The difference is negligible near the origin of AdS, where $w\simeq r$, but grows as the boundary is approached. When taking the continuum limit, the power law-divergent terms in the entropy can differ substantially, depending on which radial variable is discretized in the process. We will return to this issue in the following. 
 
The numerical calculation is performed using custom C++ code that relies on the package Eigen for linear algebra. The code works with 128-bit precision, which corresponds to 33--35 significant digits. 
 
The calculation is performed as follows: According to the discretization scheme \eqref{eq:discretization}, the degrees of freedom lie at positions $w = w_i$, with
\begin{equation}
	\frac{w_i}{a}=\frac{\pi}{2}\frac{i}{N+1},\qquad i=1,\dots , N .
\end{equation} 
We consider various radial lattices with $N = 49 + 50k$ for $k=0,\dots,10$. For all values of $N$, we compute the entanglement entropy for entangling surfaces with tortoise radii $w_n$, where $n=(k+1)j$, with $j=1,\dots , 48 $, so that the entangling surfaces have the same physical radius for all $N$. This allows for direct comparison of the entropies for various values of the UV cutoff. 

The calculation is iterated for several values of $\mu^2a^2$, both negative and positive, with $\kappa\geq 1/2$, namely
\begin{equation}
	\mu^2a^2 \in \left\{-2,-\frac{3}{2},-1,-\frac{1}{2},0,\frac{1}{2},1,\frac{3}{2},2\right\}.
\end{equation}
The values $\mu^2 a^2=0$ (massless case) and $\mu^2a^2=-2$ (conformal case) are of particular interest. 

The total entanglement entropy is given by \eqref{eq:SEE_total}. Obviously, it is impossible to perform numerical calculations for arbitrarily large $\ell$. For each of the aforementioned values of $N$ and $n$, we calculate the contribution to the entanglement entropy of the $\ell$-sectors from $\ell=0$ to $\ell=3\cdot 10^4$, and use an extrapolation in order to approximate the value of the infinite sum.

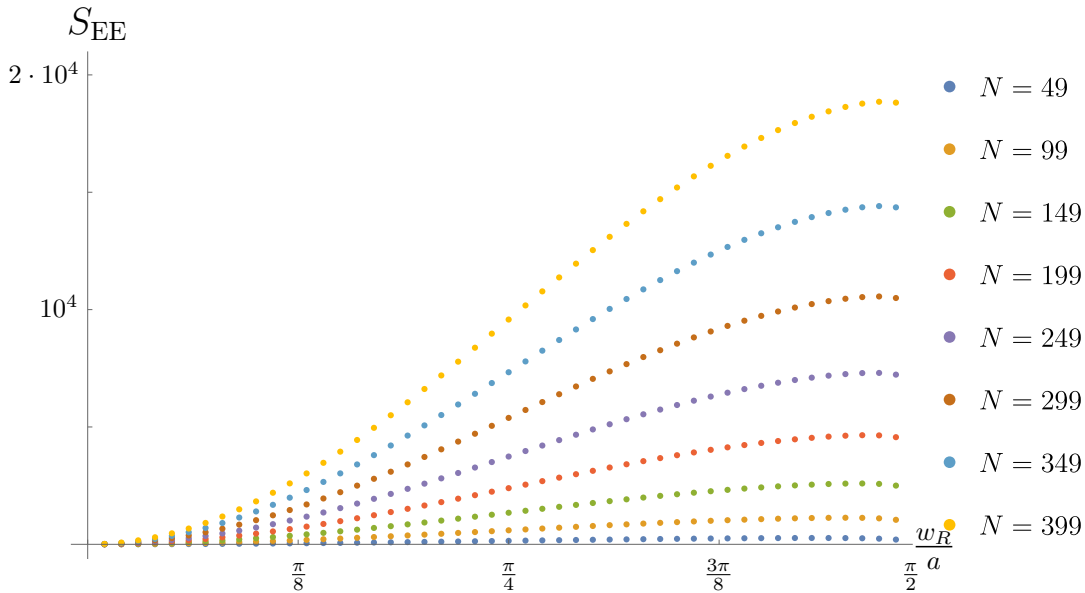
\begin{figure}[t]
	\centering
	\begin{picture}(90,51)
	    \put(0.3,43.7){{\small $2\cdot10^4$}}
	    \put(3,23.2){{\small $10^4$}}
		\put(5,2){\includegraphics[angle=0,width=0.9\textwidth]{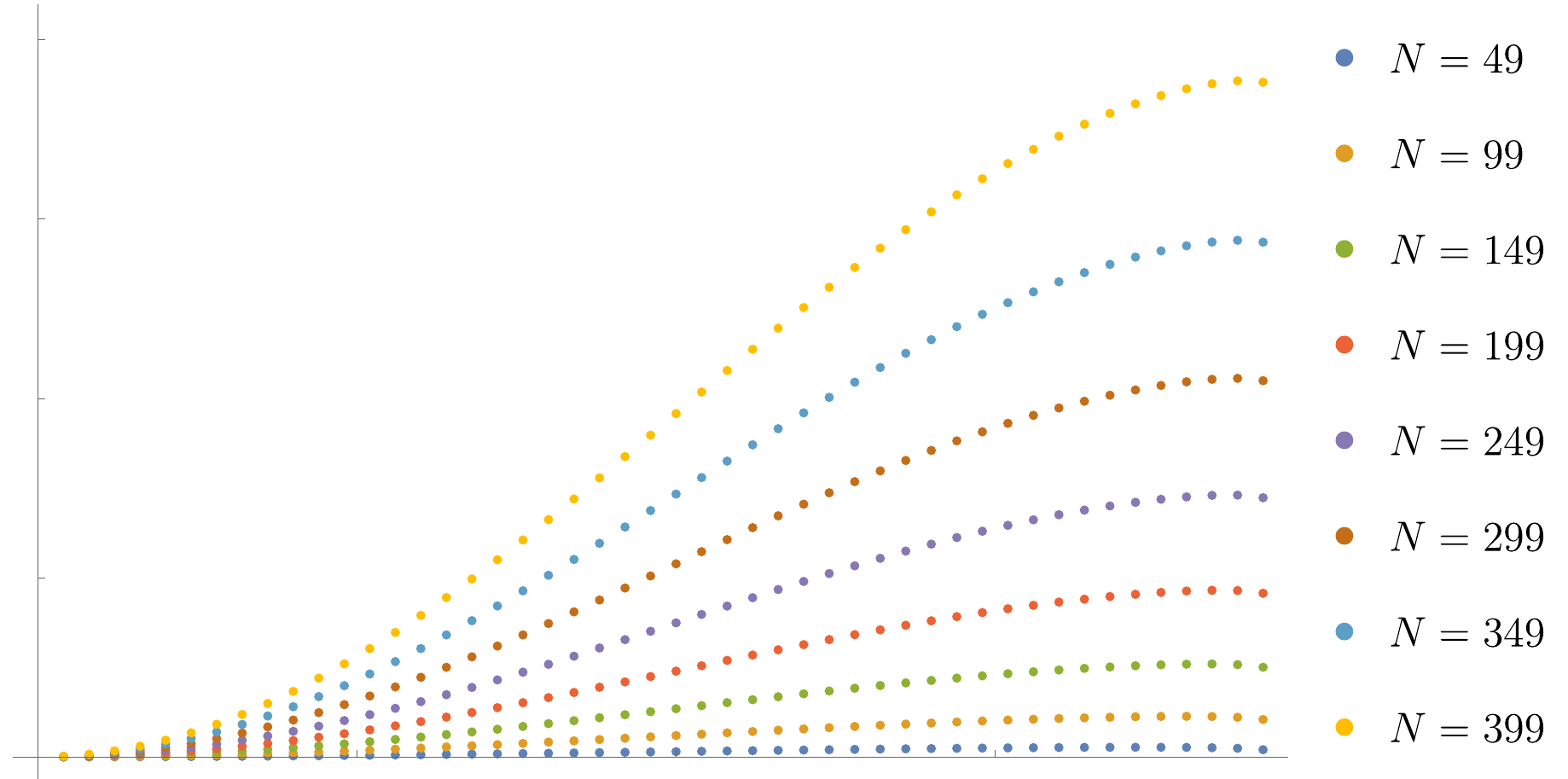}}
		\put(5.4,48){{\Large $S_\textrm{EE}$}}
		\put(24.7,0){{\small $\frac{\pi}{8}$}}
		\put(43,0){{\small $\frac{\pi}{4}$}}
		\put(60.8,0){{\small $\frac{3\pi}{8}$}}
		\put(78,0){{\small $\frac{\pi}{2}$}}
		\put(79,2.2){{\Large $\frac{w_R}{a}$}}
	\end{picture}
	\caption{Entanglement entropy as a function of the number of radial degrees of freedom $N$ and the position of the entangling surface $w_R$ for $\mu=0$. Notice that the entropy grows with $w_R$ and reaches its maximum near the boundary at $\frac{\pi}{2}$. At exactly $\frac{w_R}{a}=\frac{\pi}{2}$, it drops to 0.}
	\label{fig:plot2}
\end{figure}

Before we present the results of the numerical analysis, we discuss some qualitative features that the entanglement entropy must exhibit as a function of $N$ and the tortoise radius of the entangling surface $w_R$, in the continuous theory. Since the entropy is expected to be dominated by a divergent $1/\epsilon^2$ term, the second relation in \eqref{eq:discretization} implies that it should increase with $N^2$  when $a$ is kept constant. The entropy also grows with the area of the entangling surface, so it must reach its maximum near $w_R/a=\pi/2$. However, exactly at $\pi/2$, the entropy must vanish, as the subsystem $A$ now becomes the whole system, which is in a pure state. This peculiar behavior is a result of the finite number of degrees of freedom allowed by the discretization for any finite value of $N$. In the continuum limit, there is always an infinite number of degrees of freedom between the last entangling surface that one may consider and the boundary. As a result, the entropy will continue to grow as a function of $w_R$. For large but finite $N$, the entropy will make a sharp turn towards zero very close to the boundary. For $N\to \infty$ in the context of our scheme, the drop to zero will become vertical exactly on the boundary. This last part of the curve is unphysical, as it is a feature resulting from our particular numerical method. These expectations are verified by the numerical calculation, as can be seen in Fig.~\ref{fig:plot2}.

The first step in the analysis of the data is the determination of the value of entanglement entropy that corresponds to the infinite sum \eqref{eq:SEE_total}. This is achieved using the strategy employed in \cite{Boutivas:2024lts}. First, we calculate the truncated sum
\begin{equation}
	S_{\textrm{EE}}(n,N,\mu^2a^2;\ell_{\textrm{max}})=\sum_{\ell=0}^{\ell_{\textrm{max}}}(2\ell+1)S_{\ell m}^{\textrm{EE}}(n,N,\mu^2a^2)
\end{equation}
for several values of $\ell_\text{max}$ and study its behavior as a function of $\ell_{\textrm{max}}$. In complete analogy to \cite{Boutivas:2024lts}, see also \cite{Srednicki:1993im}, we find
\begin{multline}
	S_{\textrm{EE}}(n,N,\mu^2a^2;\ell_{\textrm{max}})	 = S_{\infty}(n,N,\mu^2a^2)\\
	+ \sum_{i=1}^{i_\textrm{max}}\frac{1}{\ell_\textrm{max}^{2i}}\left[a_i(n,N,\mu^2a^2)+b_i(n,N,\mu^2a^2)\ln \ell_\text{max}\right].
\end{multline}
By including sufficiently many subleading terms in our numerical analysis, we can extrapolate to the limit 
\begin{equation}
	S_\infty(n,N,\mu^2a^2)=\lim_{\ell_\textrm{max}\rightarrow\infty}S_{\textrm{EE}}(n,N,\mu^2a^2;\ell_\textrm{max})
\end{equation} with great accuracy.

Next, we study $S_{\infty}(n,N,\mu^2a^2)$ as a function of $\frac{a}{\epsilon}=\frac{2}{\pi}(N+1)$. It turns out that $S_{\infty}(n,N,\mu^2a^2)$ can be expanded as
\begin{equation}
	S_{\infty}(n,N,\mu^2a^2)= \frac{a^2}{\epsilon^2} S^{(2)}(n,\mu^2a^2)+ S_l^{(0)}(n,\mu^2a^2)\ln\frac{a}{\epsilon}+ S^{(0)}(n,\mu^2a^2)+R(n,N,\mu^2a^2),
\end{equation}
where the remainder $R(n,N,\mu^2a^2)$ is
\begin{equation}
	R(n,N,\mu^2a^2)=\sum_{i=1}^{i_\textrm{max}} \frac{\epsilon^i}{a^i}S^{(-i)}(n,\mu^2a^2)+\ln\frac{a}{\epsilon}\sum_{j=1}^{j_\textrm{max}} \frac{\epsilon^j}{a^j}S_l^{(-j)}(n,\mu^2a^2).
\end{equation}
The remainder vanishes for $\epsilon\rightarrow0$. The inclusion of such corrections in the expressions for the numerical fits is necessary in order to accurately calculate the terms $S^{(2)}(n,\mu^2a^2)$, $S_l^{(0)}(n,\mu^2a^2)$ and $S^{(0)}(n,\mu^2a^2)$. The first two, which pertain to the UV-divergent part of the entanglement entropy, are computed with high precision, and their behavior, as functions of the position of the entangling surface $w_R$ and the mass of the field $\mu$, is examined below. Unfortunately, a similar computation of the term $S^{(0)}(n,\mu^2a^2)$, which corresponds to the finite part of the entropy, could not lead to a definite conclusion about the functional dependence on $w_R$ and $\mu$. This is a result of the accumulation of numerical errors from the previous, dominant terms.

\subsection{Results}
\label{sec:results}

\begin{figure}[t]
	\centering
	\begin{picture}(96.5,62.5)
		\put(2,2){\includegraphics[angle=0,width=0.9\textwidth]{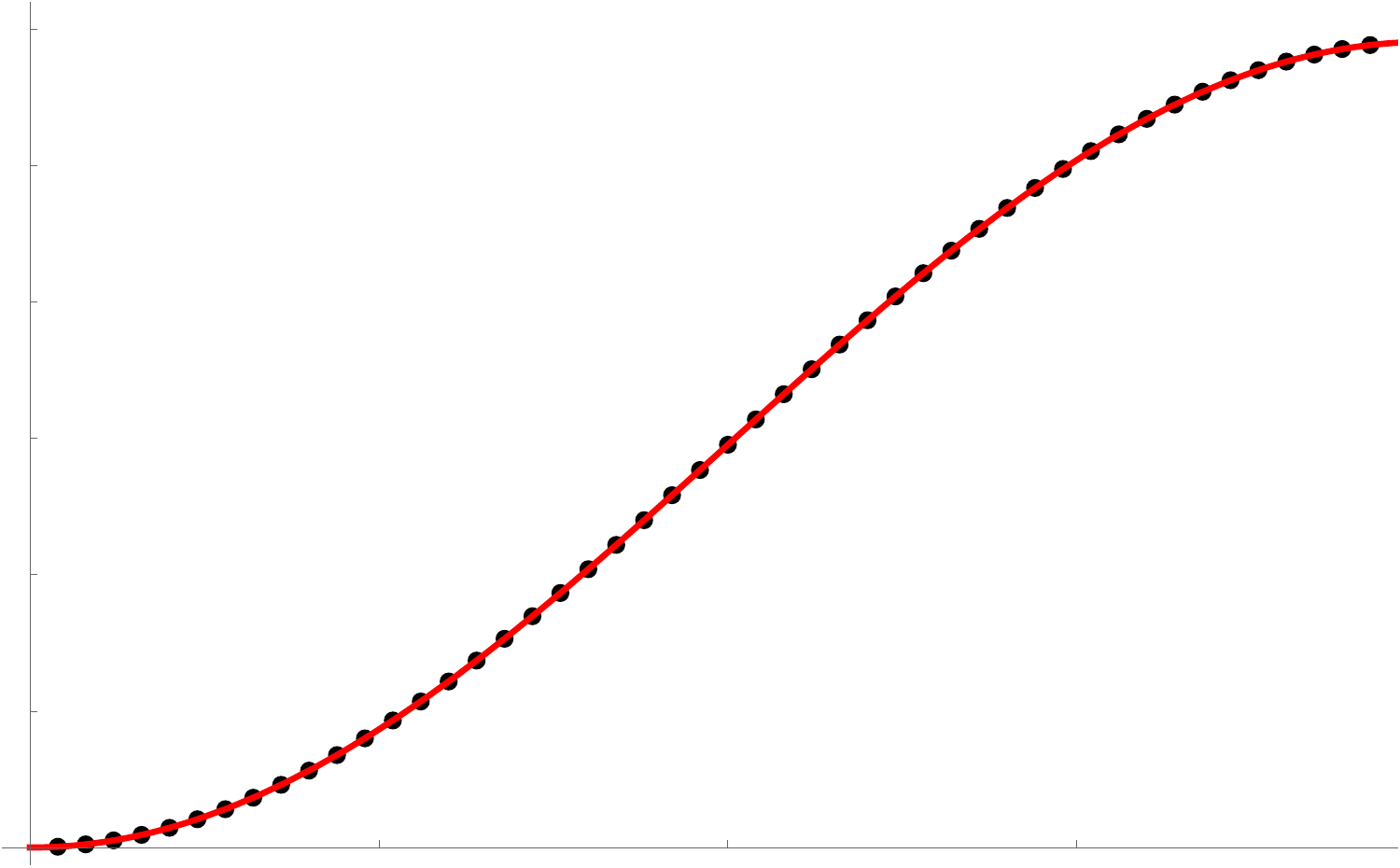}}
		\put(2.4,58.5){{\Large $S^{(2)}$}}
		\put(0,20){{\small $0.1$}}
		\put(0,37.5){{\small $0.2$}}
		\put(0,55.2){{\small $0.3$}}
		\put(25.5,0){{\small $\frac{\pi}{8}$}}
		\put(48,0){{\small $\frac{\pi}{4}$}}
		\put(70,0){{\small $\frac{3\pi}{8}$}}
		\put(91,0){{\small $\frac{\pi}{2}$}}
        \put(92.5,2.4){{\Large $\frac{w_R}{a}$}}
	\end{picture}
	\caption{Numerical fit for the determination of $S^{(2)}$ as a function of $w_R/a$ for $\mu^2a^2=0$. The data indicate that $S^{(2)}=d_1\sin^2\frac{w_R}{a}$, with $d_1\simeq0.29543145$ for all values of $\mu^2a^2$.}
	\label{fig:plot3}
\end{figure}
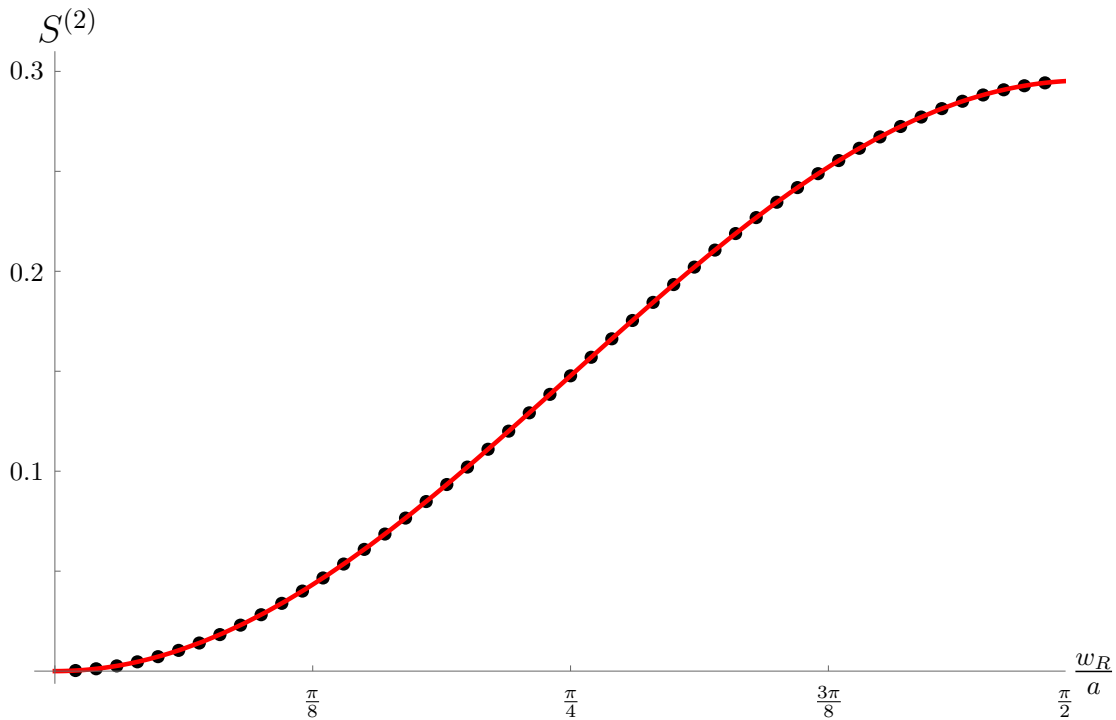

The numerical analysis of the data shows that the dominant divergent term is of the form
\begin{equation}\label{eq:square_term}
	S^{(2)}(n)=  d_1\sin^2\frac{w_R}{a},
\end{equation}
where the coefficient $d_1$ has the value
\begin{equation}\label{eq:d1_value}
	d_1\simeq 0.29543145
\end{equation}
obtained by a numerical fit. This result holds for all masses. The impressive agreement between the data and the expression \eqref{eq:square_term} is shown in Fig.~\ref{fig:plot3} for $\mu^2 a^2=0$. The plot remains the same for all values of $\mu$. For $w_R \ll a$, this expression and the numerical value of $d_1$ reproduce correctly the leading UV divergence of the entanglement entropy in flat space \cite{Srednicki:1993im}. Even though this term is regulator-dependent, the value of $d_1$ agrees with the result of \cite{Srednicki:1993im} because the entropy reduces to that in a flat background for $w_R \ll a$ and the UV regularization is the same. The same agreement is also observed in the de Sitter case for small entangling radii \cite{Boutivas:2024lts}.

For large $w_R/a$, the product $(a^2/\epsilon^2)S^{(2)}$ corresponds to the first term of \eqref{eq:adsentropy}, with the identification $A=4\pi a^2\sin^2{\frac{w_R}{a}}$. There seems to be a discrepancy, however, because the proper area of the entangling surface is $A=4\pi a^2\tan^2{\frac{w_R}{a}}$. The reason can be traced to the regularization we employed, which is based on the discretization of the radial coordinate $w/a$ that takes values in the finite interval between 0 and ${\pi}/{2}$. This differs drastically from the regularization through the discretization of the coordinate $r/a$ that takes values between 0 and infinity. The two coordinates are related through  \eqref{eq:tortoise_def}, so that a small interval of fixed proper length corresponds to $w$ or $r$ intervals related through ${\Delta w}/{\Delta r}={\cos^2{\frac{w}{a}}}$. If $r$ is discretized with a lattice spacing $\epsilon$, the number of degrees of freedom within $\Delta r$ is $\Delta r/\epsilon=(\Delta w/\epsilon)/\cos^2{\frac{w}{a}}$. This is larger by a factor of $1/\cos^2{\frac{w}{a}}$ than the number of degrees of freedom that would result from the discretization of $w$ with the same lattice spacing $\epsilon$. Notice that only the radial discretization is relevant. As far as the angular coordinates are concerned, adding the contribution of every angular momentum $\ell$ up to infinity is equivalent to a continuous distribution of degrees of freedom. The term $S^{(2)}$ quantifies the strong entanglement between the inner and outer degrees of freedom close to the entangling surface. Thus, it is expected to be proportional to the number of degrees of freedom around the entangling surface. Hence, discretizing $r$ instead of $w$ is expected to generate an additional multiplicative factor $1/\cos^2{\frac{w}{a}}$ for the leading term. This would make the result \eqref{eq:square_term} consistent with \eqref{eq:adsentropy}. At the same time, it demonstrates how extreme the regulator dependence of the entanglement entropy leading term can become. The logarithmic term we consider in the following paragraph has a coefficient that is expected to be independent of the regularization. A crucial test of the above reasoning will be provided by the form that we will deduce for this term. 

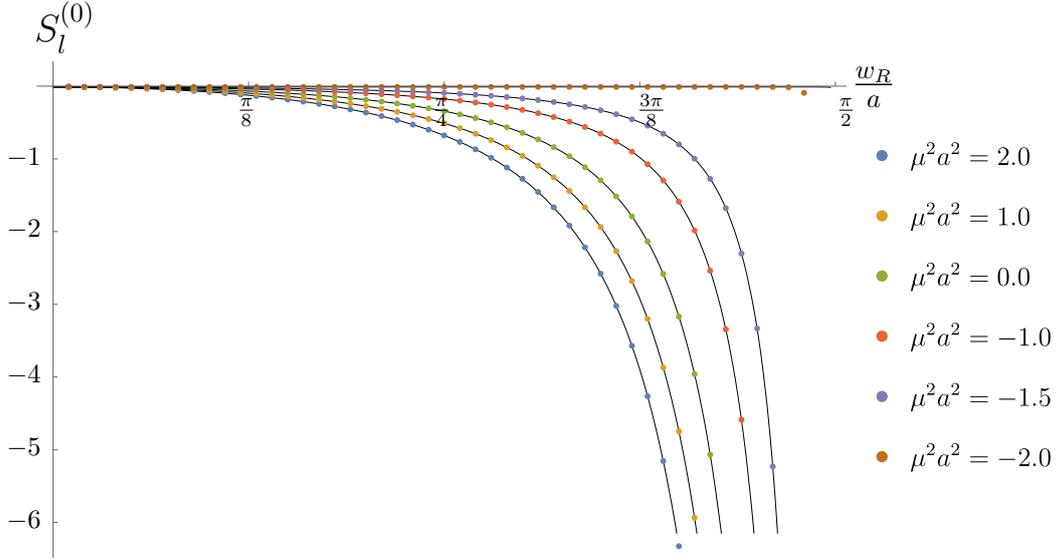
\begin{figure}[t]
	\centering
	\begin{picture}(92.3,49.5)
		\put(2.3,0){\includegraphics[angle=0,width=0.9\textwidth]{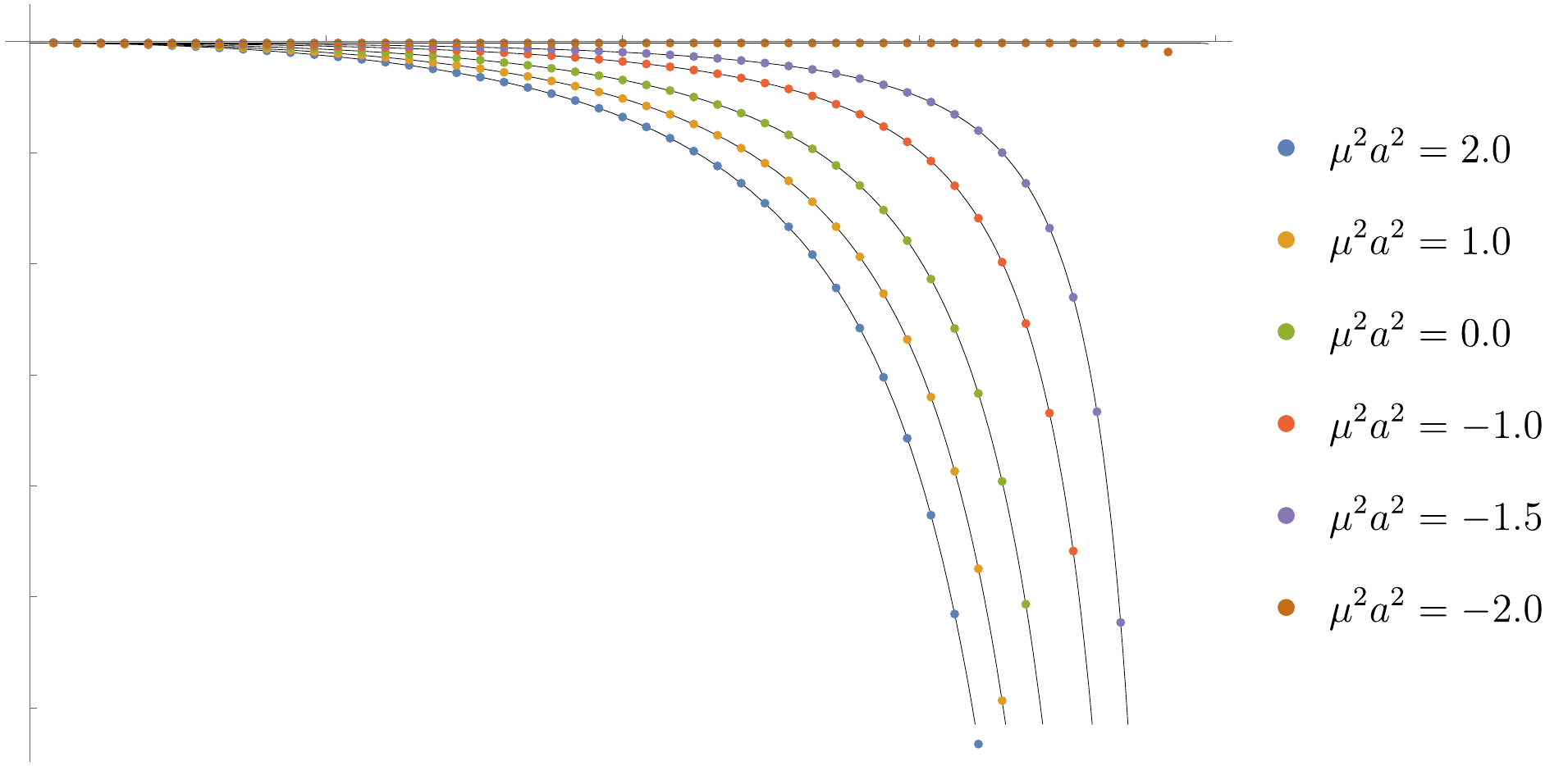}}
		\put(2.4,45.5){{\Large $S_l^{(0)}$}}
		\put(0,2.8){{\small $-6$}}
		\put(0,9.2){{\small $-5$}}
		\put(0,15.6){{\small $-4$}}
		\put(0,22.0){{\small $-3$}}
		\put(0,28.4){{\small $-2$}}
		\put(0,34.7){{\small $-1$}}
		\put(73.6,40.9){{\Large $\frac{w_R}{a}$}}
		\put(37,38.9){{\small $\frac{\pi}{4}$}}
		\put(20,38.9){{\small $\frac{\pi}{8}$}}
		\put(54.8,38.9){{\small $\frac{3\pi}{8}$}}
		\put(72.2,38.9){{\small $\frac{\pi}{2}$}}
	\end{picture}
	\caption{Numerical fits for the determination of $S^{(0)}_l$ as a function of $w_R/a$ for various values of $\mu^2a^2$. The data indicate with high precision that $S^{(0)}_l=a_l(\mu^2a^2)+b_l(\mu^2a^2)\tan^2\frac{w_R}{a}$.}
	\label{fig:plot4}
\end{figure}

Our numerical analysis shows that the coefficient of the divergent term $\ln\epsilon$ has the form
\begin{equation}
	S_l^{(0)}(n,\mu^2a^2)=a_l(\mu^2a^2) +b_l(\mu^2a^2)\tan^2\frac{w_R}{a}.
\end{equation}
The data and the numerical fits are presented in Fig.~\ref{fig:plot4} for various values of $\mu^2a^2$. Again, the agreement is impressive. It is also remarkable that the coefficient depends on $w_R$ exclusively via a term linear in $\tan^2\frac{w_R}{a}$, which is proportional to the proper area of the entangling surface. This indicates that our method indeed correctly captures the expected form of the logarithmically divergent term in equation \eqref{eq:adsentropy}. 
	
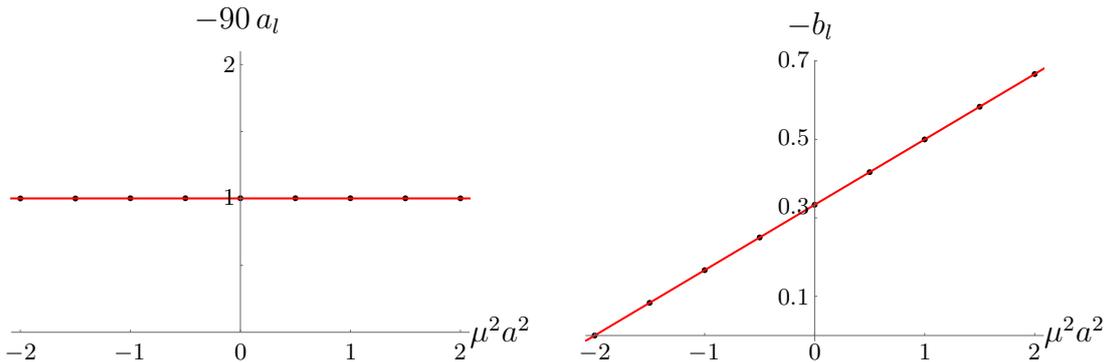
\begin{figure}[t]
	\centering
	\begin{picture}(95.5,31.5)
		\put(0,2){\includegraphics[width=0.4\textwidth]{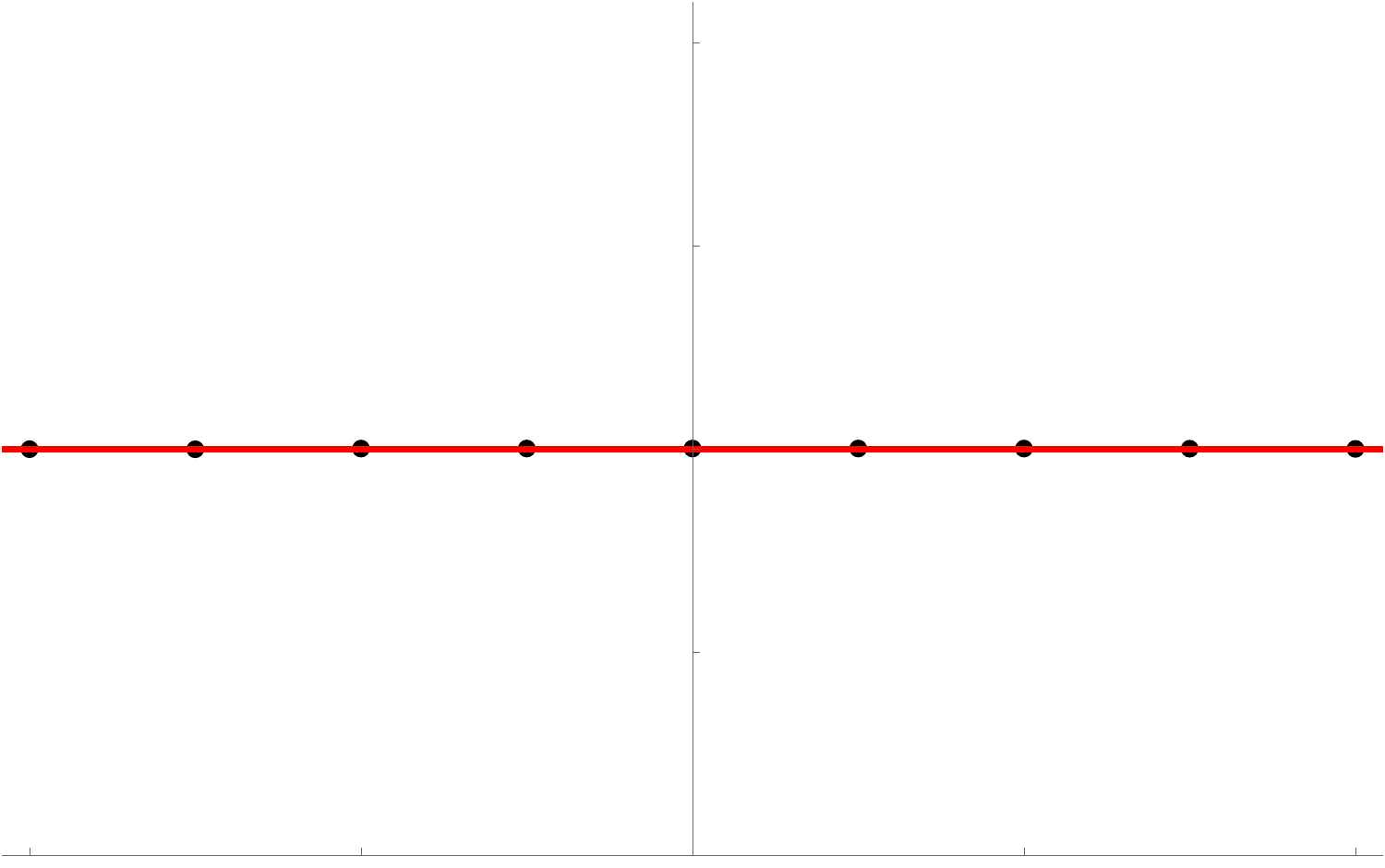} }
		\put(16,29){{\large $-90\,a_l$}}
		\put(18.5,13.5){{\footnotesize $1$}}
		\put(18.5,25.2){{\footnotesize $2$}}
		\put(40,1.5){{\large $\mu^2a^2$}}		
		\put(-0.5,0){{\footnotesize $-2$}}
		\put(9.0,0){{\footnotesize $-1$}}
		\put(19.5,0){{\footnotesize $0$}}
		\put(29.1,0){{\footnotesize $1$}}
		\put(38.6,0){{\footnotesize $2$}}						
		\put(50,1.5){\includegraphics[width=0.4\textwidth]{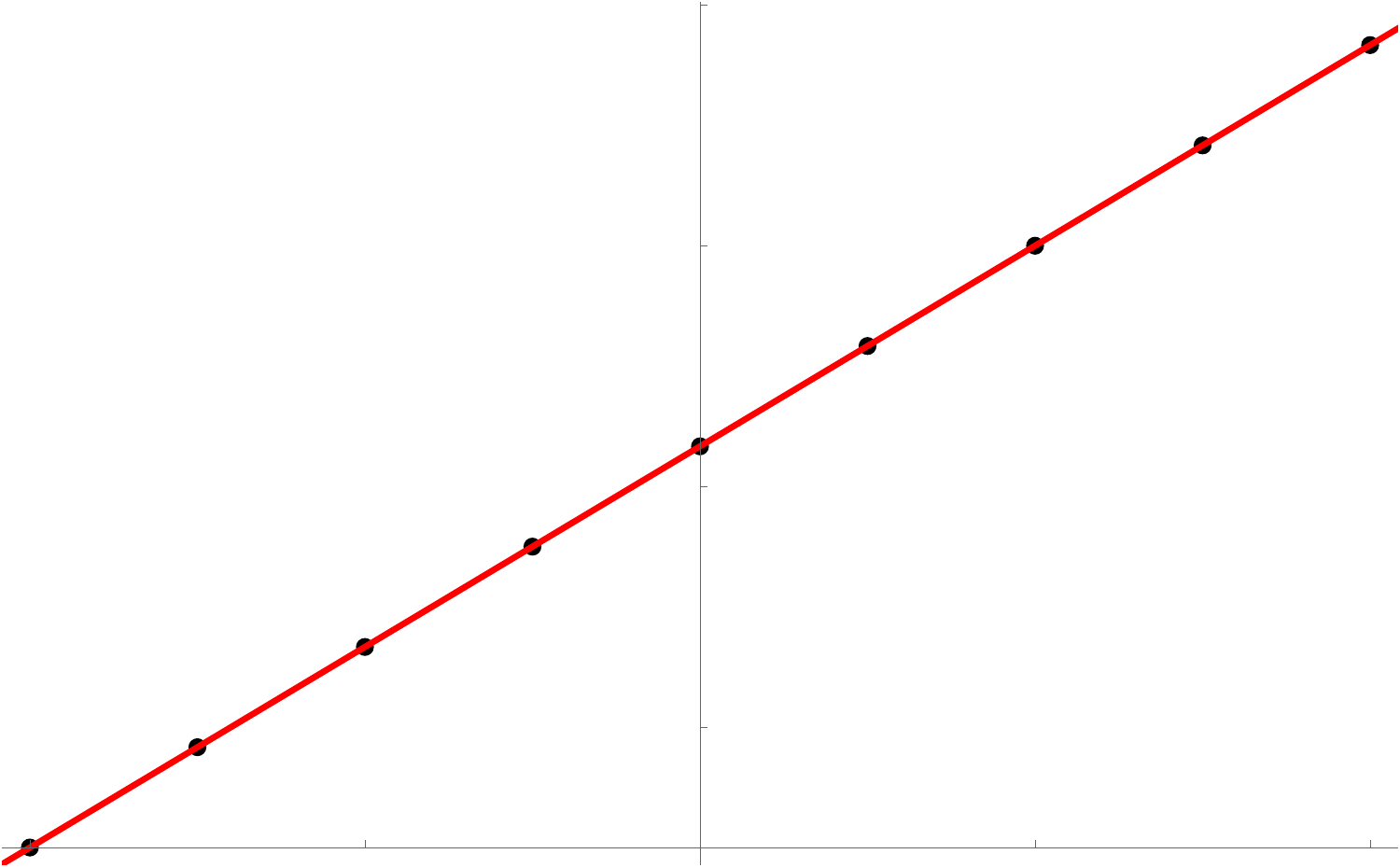} }
		\put(67.6,28.5){{\large $-b_l$}}
		\put(66.8,4.8){{\footnotesize $0.1$}}
		\put(66.8,12.7){{\footnotesize $0.3$}}
		\put(66.8,18.8){{\footnotesize $0.5$}}
		\put(66.8,25.6){{\footnotesize $0.7$}}
		\put(90,1.5){{\large $\mu^2a^2$}}
        \put(49.5,0){{\footnotesize $-2$}}
		\put(59,0){{\footnotesize $-1$}}
		\put(69.5,0){{\footnotesize $0$}}
		\put(79.1,0){{\footnotesize $1$}}
		\put(88.6,0){{\footnotesize $2$}}
	\end{picture}
	\caption{The coefficients $a_l$ and $b_l$ for various values of $\mu^2a^2$. The left plot demonstrates that $a_l=-1/90$, independently of $\mu^2a^2$, with high precision. The right plot demonstrates that $b_l$ has a linear dependence on $\mu^2a^2$ with slope equal to $-1/6.0004$ and $y-$intercept equal to $-1/3.0003$. We can confidently deduce that $b_l=-\mu^2a^2/6-1/3$.}
	\label{fig:plot5}
\end{figure}

The functions $a_l(\mu^2a^2)$ and $b_l(\mu^2a^2)$ can be determined from the data through appropriate fits, which suggest that 
\begin{align}
	a_l=&d_2={\rm const.}\label{eq:ald2} \\
	b_l(\mu^2a^2)=&d_3\,\mu^2a^2+d_4.\label{eq:bld4d5}
\end{align}
The data and the fits are displayed in Fig.~\ref{fig:plot5}. The constants are determined with an accuracy of 0.1\% to be 
\begin{equation}\label{eq:a_and_b}
	d_2=-\frac{1}{90},\qquad d_3=-\frac{1}{6},\qquad d_4= -\frac{1}{3}.
\end{equation}
They fix the form of the UV-divergent part of the entanglement entropy through \eqref{eq:adsentropy}.

\section{Discussion}
\label{sec:discussion}

We calculated the entanglement entropy of a free scalar field in AdS$_4$ space in global coordinates at its ground state, for spherical entangling surfaces centered at the origin of AdS$_4$. The calculation was performed via the original methodology of Srednicki \cite{Srednicki:1993im}. The structure of the entanglement entropy is richer than that in flat space due to the existence of another fundamental scale, the AdS length $a$. We focused on the UV-divergent terms in the entanglement entropy. Our results can be summarized in the expression 
\begin{equation}\label{eq:summary}
	S_{\textrm{AdS}} = \frac{d_1}{\epsilon^2} \sin^2 \frac{w_R}{a} + 
	\left( d_2 + \left( d_3\,\mu^2 a^2 + d_4 \right) \tan^2 \frac{w_R}{a} \right) \ln \frac{a}{\epsilon} + \textrm{finite}.
\end{equation}
The entangling sphere lies at $w = w_R$ in the tortoise coordinate system \eqref{eq:tortoise_def}. Notice that, for $a\rightarrow\infty$ with all other scales kept fixed, the tortoise coordinate $w$ becomes the radial coordinate of flat space. This allows a direct comparison with the flat-space result. The area-law term dominates the entanglement entropy. The value of its coefficient, $d_1$, which is given by \eqref{eq:d1_value},  was found to be identical to that in flat space. This result is expected, as the leading divergence arises from short-distance correlations within the UV regime of the theory, which makes it insensitive to the AdS curvature.

At first sight there seems to be a disagreement between the first term in \eqref{eq:summary} and the parameterization \eqref{eq:adsentropy} of the entropy, as the proper area of the entangling surface is $A=4\pi a^2\tan^2{\frac{w_R}{a}}$. We discussed this point in detail in subsection \ref{sec:results}, where we argued that the discrepancy is a reflection of the regulator dependence of this term that becomes extreme in our setup. There is a relative factor of $1/\cos^2{\frac{w_R}{a}}$ for the number of degrees of freedom in the discretization of the radial coordinate $r$ in the system \eqref{eq:globalr} relative to the discretization of the tortoise coordinate $w$. When this is taken into account, agreement with \eqref{eq:adsentropy} is achieved. Moreover, the correct dependence on the proper area is reproduced automatically by our analysis for the coefficient of $ \ln \frac{a}{\epsilon}$, which is independent of the choice of regulator, as can be seen through the comparison of \eqref{eq:summary} with \eqref{eq:adsentropy}.

The values of the coefficients $d_2$, $d_3$, and $d_4$ are given by \eqref{eq:a_and_b} and are consistent with the existing literature. For a conformal theory in flat space and a spherical entangling surface, the coefficient of the divergent logarithmic term in the entanglement entropy is related to the $A$-type conformal anomaly \cite{Solodukhin:2008dh}. For a scalar field, this results in a value equal to $-1/90$ \cite{Solodukhin:2008dh,Casini:2010kt,Casini:2009sr,Lohmayer:2009sq}. The agreement with the value we derived for $d_2$ is expected, as the curvature should become irrelevant for a small entangling radius, when the contributions from the terms proportional to $d_3$ and $d_4$ become negligible. The same result is obtained for a conformal theory in dS space \cite{Maldacena:2012xp,Boutivas:2024sat,Boutivas:2024lts,Abate:2024nyh}.

Our result for the coefficient $d_3$ agrees with \cite{Hertzberg:2010uv}, where it is shown that for a massive field in $(3+1)$-dimensional flat space and smooth entangling surfaces, there is a logarithmically divergent contribution to the entanglement entropy of the form
\begin{equation}\label{eq:mass_term}
	\Delta S_1=-\frac{1}{6}\mu^2 \frac{A}{4\pi}\ln\left(\mu \epsilon\right).
\end{equation}

The value of the coefficient $d_4$ can be deduced starting from the result of Solodukhin for the entanglement entropy of a conformal theory \cite{Solodukhin:2008dh}. In particular, the logarithmically divergent term for a spherical entangling surface is expected to be determined by the $A$-type conformal anomaly, which is proportional to the Euler number. This fixes the coefficient of this term to be equal to $-1/90$ for any background geometry. In the AdS case we are considering, the conformal theory would arise through a coupling ${\cal R} \phi^2/6$ of the field to gravity, which results in an effective mass term $\mu^2=-2/a^2$. The corresponding contribution to the logarithmic term must be canceled by the contribution proportional to $d_4$ in \eqref{eq:summary}. This fixes $d_4=-1/3$, which is in agreement with our result. 

Finally, we note that the coefficient $c_4=1/3$ in the dS case, appearing in \eqref{eq:SEE_expansion_mald}, is the opposite of the coefficient $d_4=-1/3$ in the AdS case. The corresponding terms in \eqref{eq:SEE_expansion_mald} and \eqref{eq:adsentropy} are related through the analytic continuation $H^2\to -1/a^2$. This is consistent with the fact that these terms must cancel the contribution from a mass term arising through the conformal coupling ${\cal R} \phi^2/6$ in both cases. 

The high accuracy with which we have computed the UV-divergent terms of the entanglement entropy indicates that it may be possible to extract information on the finite part as well. Unfortunately, this has proven to be beyond our ability without some a priori information on the structure of the terms that may appear. One particular complication is the scheme-dependence of the finite part. This is apparent in our discussion above on the dramatic difference between the regulators based on the discretization of the $w$ or $r$ radial coordinate. The study of the finite terms will be the focus of future work.

Another direction we plan to pursue in future work is the computation of the entanglement entropy in the parameter range $0\leq \kappa <1/2$, where different boundary conditions are possible. As we pointed out in subsection \ref{subsec:Discretization}, the endpoint conditions we impose are compatible with all allowed boundary conditions of the form \eqref{eq:mixed_bc} in this range of $\kappa$. It would be interesting to extend our numerical approach so that we can select any allowed boundary condition in this range of $\kappa$ in order to study the dependence of the entanglement entropy on the choice of boundary conditions. As we mentioned earlier, such an extension is challenging because one must apply an appropriate method in order to enforce a particular asymptotic form of the Hamiltonian eigenmodes, and not just their value at the boundary.

Yet another natural extension of this work is the study of entangling surfaces that are minimal surfaces anchored on the AdS boundary \cite{Sugishita:2016iel}. Such an analysis would establish a direct connection with the Ryu-Takayanagi conjecture \cite{Ryu:2006bv,Ryu:2006ef} and provide further insight into the holographic interpretation of entanglement entropy.

In addition, we compared our results with those in dS space in planar coordinates \cite{Boutivas:2024sat,Boutivas:2024lts}. However, the analytic continuation of AdS space in global coordinates is actually dS space in static coordinates. Therefore, repeating the dS calculation in this setting would also be interesting, as it would offer a direct comparison between entanglement entropies in the two geometries. In addition, performing the calculation in dS static coordinates would help clarify how entanglement entropy in this setting relates to previous results obtained in planar coordinates, shedding light on the interplay between different foliations of dS space.

\bibliographystyle{JHEP} 
\bibliography{AdS_bib}

\end{document}